\documentclass[%
 reprint,
 amsmath,amssymb,
 aps,
prb,
]{revtex4-1}

\usepackage{graphicx}%
\usepackage{dcolumn}%
\usepackage{bm}%
\newcommand*{\half}{\frac{1}{2}}
\newcommand*{\tpi}{2 \pi}
\newcommand*{\fsymco}{\frac{1}{\sqrt{\tpi}}}
\newcommand*{\lp}{\left (}
\newcommand*{\rp}{\right )}
\newcommand*{\lb}{\left [}
\newcommand*{\rb}{\right ]}
\newcommand*{\lbar}{\left |}
\newcommand*{\rbar}{\right |}
\newcommand*{\Ck}{C_{\kappa}}
\newcommand*{\Dw}{\Delta\omega}
\newcommand*{\wn}{\omega_{n}}
\newcommand*{\wm}{\omega_{m}}

\newcommand*{\Nsq}{N_\text{SQ}}
\newcommand*{\wo}{\omega_{0}}
\newcommand*{\w}{\omega}
\newcommand*{\wRT}{\omega_{\text{RT}}}
\newcommand*{\Wmod}{\Omega}
\newcommand*{\sqphase}{\theta_{\text{SQ}}}
\newcommand*{\thk}{\theta_{k}}
\newcommand*{\an}{\hat{a}_{n}}
\newcommand*{\yn}{y_{n}}
\newcommand*{\ym}{y_{m}}
\newcommand*{\am}{\hat{a}_{m}}
\newcommand*{\amdot}{\dot{\hat{a}}_{m}}
\newcommand*{\andot}{\dot{\hat{a}}_{n}}
\newcommand*{\adagn}{\hat{a}_{n}^{\dag}}
\newcommand*{\adagm}{\hat{a}_{m}^{\dag}}

\newcommand*{\bn}{\hat{b}_{n}}
\newcommand*{\bmnew}{\hat{b}_{m}}

\newcommand*{\bmdot}{\dot{\hat{b}}_{m}}
\newcommand*{\bndot}{\dot{\hat{b}}_{n}}
\newcommand*{\bdagn}{\hat{b}_{n}^{\dag}}
\newcommand*{\bdagm}{\hat{b}_{m}^{\dag}}
\newcommand*{\phid}{\hat{\phi}_{d}}
\newcommand*{\phidcl}{\phi_{d}}
\newcommand*{\phixd}{\hat{\phi}(x = d)}
\newcommand*{\phisqxd}{\hat{\phi}^{2}(x = d)}
\newcommand*{\phizn}{\phi_{n}^\text{zp}}
\newcommand*{\phizm}{\phi_{m}^\text{zp}}
\newcommand*{\phizmsq}{\lp\phi_{m}^\text{zp}\rp^2}
\newcommand*{\bigD}{\mathcal{D}}

\newcommand*{\EJ}{E_{J}}
\newcommand*{\EJo}{E_{J0}}
\newcommand*{\df}{\delta f}
\newcommand*{\dsq}{d_{sq}}
\newcommand*{\Phio}{\Phi_{0}}
\newcommand*{\rphio}{\varphi_{0}}
\newcommand*{\kapn}{\kappa_{n}}
\newcommand*{\kapm}{\kappa_{m}}
\newcommand*{\Deln}{\Delta_{n}}
\newcommand*{\Delm}{\Delta_{m}}
\newcommand*{\bigJ}{\mathcal{J}}
\newcommand*{\dsqdef}{\left \langle \lp \frac{E_{J1} - E_{J2}}{E_{J1} + E_{J2}} \rp^2 \right \rangle}
\newcommand*{\betan}{\beta_{n}}

\newcommand*{\betam}{\beta_{m}}
\newcommand*{\betadotm}{\dot{\beta}_{m}}
\newcommand*{\betaout}{\underline{\beta^{\text{out}}}}
\newcommand*{\betaoutm}{\beta^{\text{out}}_{m}}
\newcommand*{\betain}{\underline{\beta^{\text{in}}}}

\newcommand*{\ofw}{[\omega]}

\newcommand*{\kapmat}{\boldsymbol{\kappa}}
\newcommand*{\kapemat}{\boldsymbol{\kappa^{e}}}
\newcommand*{\bpA}{(\textbf{A}) }
\newcommand*{\bpB}{(\textbf{B}) }
\newcommand*{\bpC}{(\textbf{C}) }
\newcommand*{\bpD}{(\textbf{D}) }
\newcommand*{\bpE}{(\textbf{E}) }
\newcommand*{\bpF}{(\textbf{F}) }
\newcommand*{\bpG}{(\textbf{G}) }

\newcommand*{\mus}{\mu \text{s}}
\newcommand*{\pdt}{\partial_{t}}
\newcommand*{\sinc}{\text{sinc}}
\newcommand*{\no}{n_{0}}
\newcommand*{\km}{k_{m}}

\newcommand*{\thmod}{\theta_{\text{mod}}}
\newcommand*{\thawg}{\theta^{\text{AWG}}}
\newcommand*{\kf}{k_{f}}
\newcommand*{\Tout}{T^{\text{out}}}

\newcommand*{\un}{\underline{u_{n}}}
\newcommand*{\udagn}{\underline{u^{\dag}_{n}}}
\newcommand*{\um}{\underline{u_{m}}}

\newcommand*{\Un}{\mathcal{U}_{n}}
\newcommand*{\Unp}{\mathcal{U}_{n+1}}
\newcommand*{\sgn}{\text{sgn}}
\newcommand*{\Zm}{Z_{m}}
\newcommand*{\Zo}{Z_{0}}
\newcommand*{\lgrngn}{\mathcal{L}}
\newcommand*{\Cn}{C_{n}}
\newcommand*{\EJn}{E_{Jn}}
\newcommand*{\LJn}{L_{Jn}}
\newcommand*{\LJo}{L_{J0}}

\newcommand*{\phiddot}{\ddot{\phi}}
\newcommand*{\phidotn}{\dot{\phi}_{n}}
\newcommand*{\phin}{\phi_{n}}
\newcommand*{\etan}{\eta_{n}}
\newcommand*{\etam}{\eta_{m}}

\newcommand*{\etaon}{\eta_{0n}}

\newcommand*{\dn}{d_{n}}
\newcommand*{\xn}{x_{n}}
\newcommand*{\zn}{z_{n}}
\newcommand*{\lam}{\lambda}
\newcommand*{\eps}{\epsilon}
\newcommand*{\thplus}{\Theta_{+}}
\newcommand*{\intallreal}{\int_{-\infty}^{\infty}}
\newcommand*{\wnex}{\wn^\text{ex}}

\newcommand*{\wmread}{\wm^{\text{read}}}
\newcommand*{\Gnin}{G_{n}^{\text{in}}}
\newcommand*{\Gnout}{G_{n}^{\text{out}}}
\newcommand*{\Gmin}{G_{m}^{\text{in}}}
\newcommand*{\Gmout}{G_{m}^{\text{out}}}
\newcommand*{\vo}{v_{0}}
\newcommand*{\ki}{k_{\text{i}}}
\newcommand*{\kr}{k_{\text{r}}}
\newcommand*{\kt}{k_{\text{t}}}
\newcommand*{\thJ}{\theta_{J}}
\newcommand*{\Ar}{A_{\text{r}}}
\newcommand*{\At}{A_{\text{t}}}
\newcommand*{\betaul}{\underline{\beta}}
\newcommand*{\keff}{k_\text{eff}}
\newcommand*{\Hhat}{\hat{H}}
\newcommand*{\Hbold}{\boldsymbol{H}}
\newcommand*{\Sbold}{\boldsymbol{S}}

\begin{document}

\preprint{APS/123-QED}

\title{Electric fields for light: \\ Propagation of microwave photons along a synthetic dimension}

\author{Nathan R. A. Lee}
 \email{nlee92@stanford.edu} 
\author{Marek Pechal} 
\author{E. Alex Wollack} 
\author{Patricio Arrangoiz-Arriola} 
\author{Zhaoyou Wang}
\author{Amir H. Safavi-Naeini}
 \email{safavi@stanford.edu} 
\affiliation{%
 Department of Applied Physics and Ginzton Laboratory, Stanford University\\
 348 Via Pueblo Mall, Stanford, California 94305, USA
}%

\date{\today}

\begin{abstract}
The evenly-spaced modes of an electromagnetic resonator are coupled to each other by appropriate time-modulation, leading to dynamics analogous to those of particles hopping between different sites of a lattice. This substitution of a real spatial dimension of a lattice with a ``synthetic'' dimension in frequency space greatly reduces the hardware complexity of an analog quantum simulator. Complex control and read-out of a highly multi-moded structure can thus be accomplished with very few physical control lines. We demonstrate this concept with microwave photons in a superconducting transmission line resonator by modulating the system parameters at frequencies near the resonator's free spectral range and observing propagation of photon wavepackets in time domain. The linear propagation dynamics are equivalent to a tight-binding model, which we probe by measuring scattering parameters between frequency sites. We extract an approximate tight-binding dispersion relation for the synthetic lattice and initialize photon wavepackets with well-defined quasimomenta and group velocities. As an example application of this platform in simulating a physical system, we demonstrate Bloch oscillations associated with a particle in a periodic potential and subject to a constant external field. The simulated field strongly affects the photon dynamics despite photons having zero charge. Our observation of photon dynamics along a synthetic frequency dimension generalizes immediately to topological photonics and single-photon power levels, and expands the range of physical systems addressable by quantum simulation.

\end{abstract}

\maketitle

\section{\label{sec:intro} Introduction}

Light does not have charge. Accordingly, electric and magnetic fields do not strongly affect the propagation of photons. Moreover, photons interact very weakly with one another. Because of this, experimental studies of the dynamics of many photons have lacked the rich complexity found in the many-body quantum physics of condensed matter. Recent progress has led to the development of photonic-matter devices that imbue light with the properties of matter, including mass, charge, and many-body interactions \cite{Koch2010, Ma2017, Jia2018}. These devices have attracted increasing attention over the past decade for their ability to simulate topological and disordered many-body physics with single-site resolution, which is difficult to measure for individual atoms or electrons in a condensed-matter system \cite{Ozawa2019}. A common first step in realizing photonic matter is engineering a lattice Hamiltonian that allows photons to propagate between sites; the lattice may include both spatial dimensions (such as islands of superconducting metal or atomic positions in an optical trap) and synthetic dimensions along internal degrees of freedom (such as modes of a resonator or spin states of an atom). Synthetic dimensions allow the observation of higher-dimensional physics in structures with fewer spatial dimensions which are often easier to construct \cite{Yuan2018}, motivating theoretical proposals across a wide variety of physical systems \cite{Tsomokos2010, Fang2012, Peropadre2013, Schmidt2015, Luo2015, Luo2017, Ozawa2016, Price2017, Yuan2016, Yuan2016a, Sundar2018, Anderson2016, Yang2016, Ostmann2018}. While synthetic dimensions have been studied experimentally in ultracold atoms \cite{Mancini2015, Stuhl2015, Livi2016, Meier2017}, fiber-optical systems \cite{Bell2017, Qin2018, Dutt2019} and optical waveguide arrays \cite{Regensburger2012, Lustig2019}, they have received limited experimental attention in superconducting circuits \cite{Zakka-Bajjani2011, Chang2018} despite the rapid development of circuit quantum electrodynamics (“circuit QED”) \cite{Vool2017} as a platform for quantum science. Lattice simulation experiments in circuit QED have focused  on spatial dimensions defined by multiple transmon qubits \cite{Roushan2017, Reagor2018, Ma2019}, coplanar waveguide resonators \cite{Underwood2012, Fitzpatrick2017, Kollar2019} or other microwave cavities \cite{Owens2018}, requiring either a large number of control lines or an inability to directly access the internal lattice sites. Moreover, as the dimensionality of the problems increase, synthetic lattices become an attractive route to realizing full control and read-out in a hardware-efficient way, as they allow for nearly arbitrary connectivity~\cite{Naik2018,Pechal2019,Hann2019}. %
\begin{figure}[b]
\includegraphics{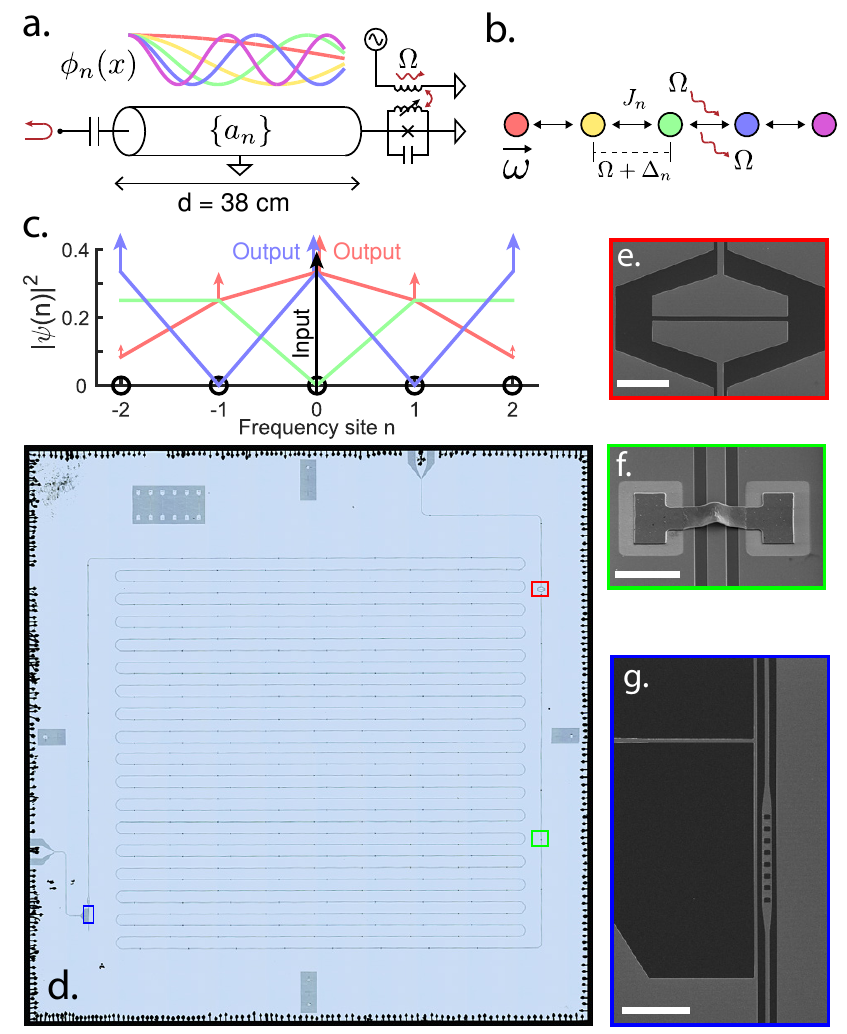}
\caption{\label{fig:circuit_diagram} \textbf{Coplanar waveguide resonator with modulated SQUID termination}. \bpA Schematic of resonator including external coupling and flux line. Standing waves depict the local amplitude of 5 adjacent modes; the lowest 5 modes are displayed for visual clarity though not directly accessed in this work. \bpB Schematic of lattice sites along the synthetic dimension when the flux bias is modulated at frequency $\Wmod$ near the average free spectral range. Coupling to the next higher (lower) frequency site can be envisioned as absorbing (emitting) a photon from (to) the modulating field. \bpC Three lowest-energy eigenfunctions of a 5-site tight-binding Hamiltonian with uniform nearest-neighbor coupling. The narrow-band input signal shown excites only eigenfunctions with nonzero amplitude at site $0$ (red and blue), and subsequent output signals at other sites are emitted only from the red and blue eigenfunctions. \bpD Optical micrograph of the resonator, fabricated with aluminum on a 14x14mm chip of high-resistivity silicon. The device layout is similar to Refs. \cite{Sundaresan2015, Zhong2019}. Key features are shown in scanning electron micrographs: \bpE external-coupling capacitor, \bpF superconducting airbridges to suppress parasitic slotline resonances, and \bpG SQUID array termination and flux line. The scale bars are respectively 50, 20, and 50 microns. The inward crimping of the airbridge span in \bpF is a systematic fabrication defect and is discussed in the Appendix.}
\end{figure}

We experimentally demonstrate an approach to realizing a synthetic lattice in the modes of a multi-mode superconducting resonator and observe coupled-mode dynamics of photons under parametric modulation. The resonator, depicted schematically in Fig. \ref{fig:circuit_diagram}, is a coplanar waveguide (CPW) of length $d \approx 38 \text{ cm}$ terminated at one end by an array of $\Nsq = 8$ superconducting interference devices (SQUIDs) that functions approximately as an $LC$ circuit with tunable inductance. The resonant frequencies within the experimentally-accessible band (4-8 GHz) are nearly equally-spaced by $\Dw / \tpi \approx 155 \text{ MHz}$ according to the condition for round-trip constructive interference given by:
\begin{equation} \label{eq:round_trip_modes}
    \frac{\wn}{\wRT} = n + \sqphase[\wn]/\tpi
\end{equation}
Here $\sqphase$ is the phase of the reflection coefficient off the $LC$ equivalent circuit into a CPW with constant wave impedance $\Zo = \sqrt{l/c}$ \cite{Pozar2012}, and $\wRT \equiv \pi v_p / d \approx \Dw$ is the round-trip frequency at constant phase velocity $v_p = 1/\sqrt{lc}$. $\sqphase[\omega] \in (-\pi, \pi)$ increases monotonically with $\omega$ \cite{Foster1924} and defines the tuning range of each resonance. External coupling occurs at the end opposite the SQUID array through a capacitance $\Ck \approx 15 \text{ fF}$, designed to achieve near-critical coupling to modes between 4 GHz and 6 GHz. Each resonant frequency $\wn/\tpi$ can be tuned over a $17 \text{ MHz}$ range using an on-chip flux line, enabling simultaneous parametric modulation of all resonant frequencies. As we will see below, modulation at frequencies $\Wmod = m \Dw + \delta$ near integer multiples of the average free spectral range introduces tight-binding coupling between $m^{th}$ nearest-neighbor modes, and in the lossless limit the resonances of the modulated Hamiltonian form a Floquet quasi-energy spectrum given by $\omega_{k, l} = \omega_k + l \Wmod$ \cite{Gomez-Leon2013, Dutt2019}.

\section{Tight-binding model from parametric modulation}

\subsection{\label{sec:matrix_elements} Expressions for the matrix elements}

We describe the modulated Hamiltonian with a formalism previously used to describe parametric amplification in tunable superconducting cavities \cite{Wallquist2006, Wustmann2013, Wustmann2017}. Setting $\hbar = 1$ unless otherwise indicated, the Hamiltonian of the waveguide resonator is given by
\begin{eqnarray} \label{eq:basic_modulated_H}
    \Hhat(t) = \sum_n \wn \adagn \an + \bigD(t) \phisqxd
\end{eqnarray}
where $\an$ ($\adagn$) is the annihilation (creation) operator for photons in mode number $n$, $\phixd \equiv \phid$ is the phase difference across the SQUID array, $\bigD(t)$ is a periodic drive with fundamental frequency $\Wmod$, and each $\wn$ is fixed to its value when $\bigD(t) = 0$. The interaction term $\phid^2 = 
\sum_{mn} \phizm \phizn (\am + \adagm)(\an + \adagn)$ is the modulated inductive energy in the SQUID array arising from a quadratic approximation to its Hamiltonian which is proportional to ${\cos(\phi_{d})}$, and where we neglect the quartic and higher-order terms at sufficiently low mean photon numbers.
Given an $\Omega-$periodic modulation $\bigD(t) = \sum_{k=0}^{\infty} \bigD[k] \cos(k\Wmod t + \thk)$ and choosing $\wo$ as the center frequencies of one of the modes, we write input-output equations for the mode spaced $n$ lattice sites away from the one at $\wo$, in a rotating frame as $\an(t) = \bn(t)e^{-i(\wo + n \Wmod)t}$:
\begin{multline} \label{eq:rot_frame_evol}
    \bmdot = \lp -i \Delm - \frac{\kapm}{2} \rp \bmnew  ...
    \\
    - i \phizm  \bigD(t) \sum_n  \phizn \lp \bn e^{-i (n-m)\Wmod t} +  \bdagn e^{i (n+m) \Wmod t}\rp
    \\
    + \sqrt{\kapm^{e}} \bmnew^\text{in} e^{-i(\wm^\text{in} - \wo - m \Wmod )t}
\end{multline}
with the boundary condition
\begin{equation} \label{eq:rot_frame_output}
    \bmnew^\text{out} =  \bmnew^\text{in} - \sqrt{\kapm^{e}} \bmnew e^{i (\wm^\text{in} - \wo - m \Wmod)t}.
\end{equation}
 Here $\kapm = \kapm^{e} + \kapm^{i}$ is the sum of loss rates to measurement coupling (e) and internal degrees of freedom (i). The rotating-frame frequencies are equally-spaced at the modulation frequency, and we choose $\wo$ to be the center frequency of the resonator's $32^\text{nd}$ harmonic at $4.989~\text{GHz}$, for reasons outlined in sections~\ref{sec:measuring_s_params} and~\ref{sec:resonant_mod_dynamics}.  The input frequencies $\wm^\text{in}$ can be set equal to the comb frequencies such that $\wm^\text{in} - \wo - m \Wmod = 0$. We next expand $\bigD(t) = \sum_{k=0}^{\infty} \bigD[k] \cos(k\Wmod t + \thk)$, and use the rotating wave approximation to remove all non-constant terms in the Hamiltonian \cite{Hammerer2014}; this requires $\Wmod \gg (\Delm, \kapm, \phizm\phizn \, \bigD[k])$ which is satisfied for all $m$ accessed in this work:
 \begin{multline} \label{eq:time_domain_rot_frame}
     \bmdot = \lp -i \Delm - \frac{\kapm}{2} \rp \bmnew - i\sum_{k} J_{m, m + k} \hat{b}_{m + k} + \sqrt{\kapm^{e}} \bmnew^\text{in}
     \\
     = -i \sum_{n} \lp \Hbold_{mn} - i \frac{\kapm}{2} \delta_{mn} \rp \bn + \sqrt{\kapm^{e}} \bmnew^\text{in} 
 \end{multline}
The elements of the matrix $\Hbold$ contain the on-site energies $\Delta_m$ as well as the $k^\text{th}$ nearest-neighbor couplings $J_{m, m+k}$. We focus here on nearest-neighbor coupling ($k = 1$), and investigate more general couplings in the Appendix. In the limit where all junctions in the SQUID array are identical, the nearest-neighbor coupling rates can be expressed as
\begin{equation}
    J_{m, m+1} = -\EJo \, \phizm \phi^{zp}_{m+1} \sin(F) \bigJ_{1}(\df) e^{-i \theta_{1}},
\end{equation}
where $\EJo$ is the maximum Josephson energy of the array, $\bigJ_1$ is a Bessel function of the first kind, and the modulated flux threading each SQUID is given by $\frac{\pi}{\Phi_0} \Phi(t) = F + \df \cos(\Wmod t + \theta_1)$ (see Appendix \ref{app:rot_frame}).

\begin{figure*}[t]
\includegraphics{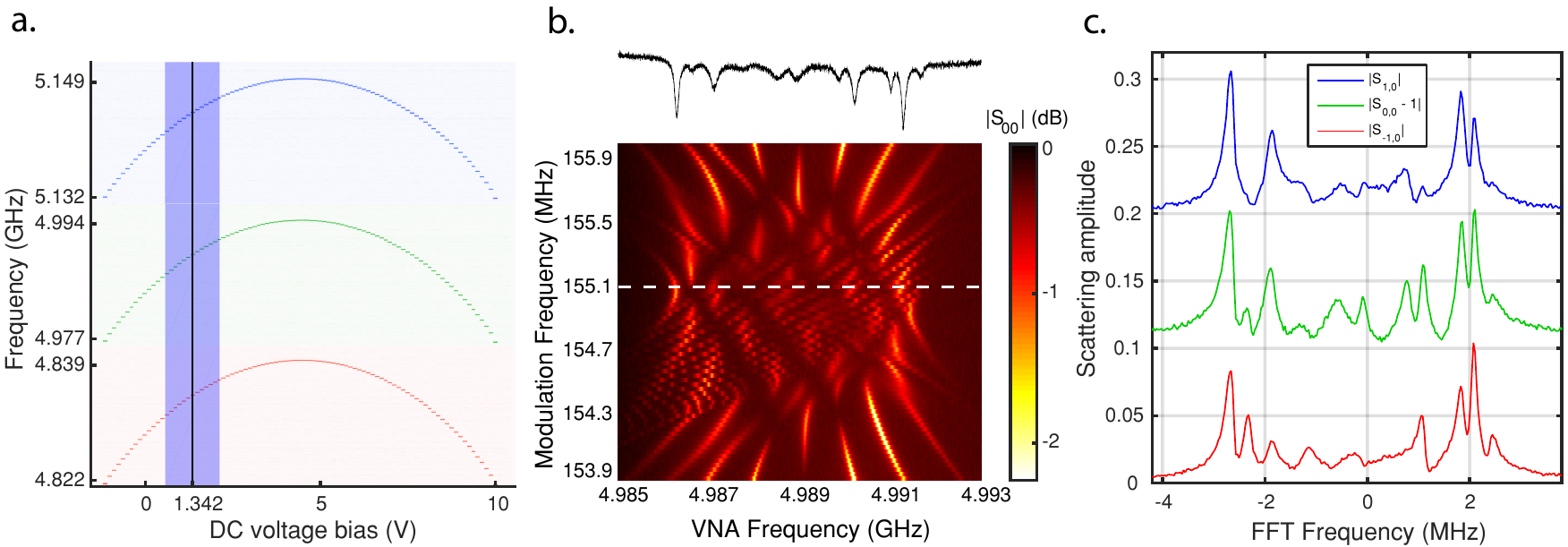}
\caption{\label{fig:spectra} \textbf{Experimental spectra demonstrating flux tunability and tight-binding coupling}. \bpA Parallel tuning of modes 31, 32 and 33 ($n = -1, 0, 1$) over 17 MHz as the flux line voltage bias is swept (the frequency axis is collapsed between each mode). The vertical bar denotes the DC bias used for modulation experiments; the blue overlay displays a typical flux modulation amplitude equivalent to $0.062 \Phio$. \bpB Continuous-wave (CW) reflection spectra at site $n = 0$, measured as a VNA reflection coefficient as the modulation frequency is swept from $153.9$ to $156~\text{MHz}$. VNA spectra correspond to eqn. \ref{eq:scattering_with_input_on}. The dashed line and corresponding line trace at $155.1~\text{MHz}$ correspond to ``resonant modulation'' at $\Wmod/\tpi =  155.1~\text{MHz}$ used in subsequent experiments. \bpC Transient spectra for scattering between input site $n = 0$ and output sites $(-1, 0, 1)$, corresponding to eqn. \ref{eq:scattering_with_input_off}. Traces for $\Sbold_{00}$ (green) and $\Sbold_{10}$ (blue) are vertically shifted by $0.1$ and $0.2$ respectively. The green trace may be compared to the VNA trace in \bpB as they represent the same scattering parameter with \bpB and without \bpC direct reflection.
}
\end{figure*}

\subsection{\label{sec:flux_tunability} Characterizing flux tunability}
The superconducting circuit is shielded from external oscillating fields and cooled to $T_\text{base} \lesssim 10~\text{mK}$ at the base plate of a dilution refrigerator; the full experimental setup is discussed in Appendix \ref{app:exp_details}. Before performing a flux-modulation experiment, the flux line must be calibrated to offset stray DC magnetic fields and extract the mutual inductance between the flux line and SQUID loops, which then provides an estimate of the parametric modulation amplitude.

We perform single-port reflection measurements using a vector network analyzer (VNA) to locate resonant frequencies and measure their tunability. We bias the flux line with an isolated DC source across an RC lowpass filter with $R_\text{tot} \approx 4 \, \text{k}\Omega$, which determines the current sent to the chip. To extract approximate tuning parameters, we sweep the bias voltage over several volts and fit the extracted resonant frequencies to the following model \cite{Koch2007, Wustmann2013, Eichler13}, which is equivalent to eqn. \ref{eq:round_trip_modes}:
\begin{multline} \label{eq:tuning_model}
    \tan(\yn) + \frac{\yn}{B} = ...
    \\
    \frac{A}{\yn} \sqrt{\cos^2[G (V - V_{ss})] + \dsq^2 \sin^2[G (V - V_{ss})]}
\end{multline}
where
\begin{equation} \label{eq:yn_def}
    \yn \equiv \frac{\pi \wn}{\wRT}.
\end{equation}
Fitted parameters using a least-squares cost function are given in Table \ref{tab:tuning_fit_params}. We find empirically that values $B$ and $\dsq^2$ are not well-constrained by this fitting method in the sense that the cost function takes similar values for a wide range of ($B$, $\dsq^2$) with only small deviations in other parameters. However the primary experimental goal is to calibrate $G$, the voltage sensitivity, and $V_{ss}$, the voltage offset, which \textit{are} well-constrained by the period and phase of the tuning function. An example of parallel flux-tuning for three modes is given in Fig. \ref{fig:spectra}a.

\begin{table}[b]
\caption{\label{tab:tuning_fit_params}%
\textbf{Fit parameters extracted from flux bias tuning}, with equivalent expressions in terms of circuit parameters. $M$ is the mutual inductance between the on-chip flux line and a single SQUID loop, and $L_{s0} = \rphio^2/\EJo$ is the minimum Josephson inductance of the SQUID array. The average defining $\dsq^2$ is taken over all SQUIDs in the array; the distribution for this average is given in the Appendix. $\omega_s/\tpi$ is the Josephson plasma frequency of the SQUID array and provides an upper bound for modulation frequencies. Although not explicitly included in the tuning model, the Josephson plasma frequency is related by $\omega_s = \wRT \sqrt{AB}/\pi$.
}
\begin{ruledtabular}
\begin{tabular}{lcr}
\textrm{Fit parameter}&
\textrm{Equiv. expression}&
\textrm{Value}\\
\colrule
$G/\pi$ & $M/\Phio R_{tot}$ & $0.0796$ V$^{-1}$\\
$V_{ss}$ & $-\pi \Phi_{stray} / G \Phio$ & $4.481$ V\\
$\wRT/\tpi$ & $v_p/2d$ & $155.52$ MHz\\
$A$ & $ld / L_{s0}$ & $40.11$ \\
$B$ & $cd / C_{s}$ & $4479$ \\
$\dsq^2$ & $\dsqdef$  & $ <0.01$ \\
$\omega_{s} / \tpi$ & $(\tpi\sqrt{L_{s0} C_{s}})^{-1}$ & $21.0$ GHz \\
\end{tabular}
\end{ruledtabular}
\end{table}

\subsection{\label{sec:measuring_s_params} Measuring scattering parameters of the lattice}
We measure the spectrum of our device using two techniques. First, we drive the resonator and flux line with continuous signals and measure the steady-state voltage reflected from the resonator to obtain reflection coefficients. Next, we drive the resonator and flux line with pulses and measure the emitted voltage over time. For large coherent states in each site ($|\beta|^2\approx 10$ to $1000$ in experiments), we replace the operators $\bmnew$ with their classical averages $\betam$ and solve the input-output equations in the Fourier domain. We measure emitted voltages both with (Eqn. \ref{eq:scattering_with_input_on}) and without (Eqn. \ref{eq:scattering_with_input_off}) the directly-reflected input field:
\begin{gather}
    \label{eq:scattering_with_input_on}
    \betaout \ofw = \lb \boldsymbol{I} + i \sqrt{\kapemat} \lp \Hbold - i \frac{\kapmat}{2} - \omega \boldsymbol{I}  \rp^{-1} \sqrt{\kapemat}\rb \betain \ofw \\
    \label{eq:scattering_with_input_off}
    \betaout \ofw = i \sqrt{\kapemat} \lp \Hbold - i \frac{\kapmat}{2} - \omega \boldsymbol{I} \rp^{-1} \sqrt{\kapemat} \underline{\tilde{\beta}}(t = 0^+)
\end{gather}
where $\betaul \equiv (\beta_0, ...\, , \beta_m, ...\,)^{T}$ are vectors of coherent-state amplitudes and $\boldsymbol{\kappa^{(e)}} \equiv \text{diag}(\{ \kapm^{(e)} \})$. Eqn. \ref{eq:scattering_with_input_on} describes a steady-state scattering matrix $\Sbold_{\beta \alpha}\ofw$ between lattice sites, whose diagonal elements can be measured experimentally with a standard VNA. Eqn. \ref{eq:scattering_with_input_off} describes transient scattering following the preparation of a known initial state $\underline{\tilde{\beta}}(t = 0^+) = \lp \sqrt{\kapemat}\rp^{-1}\beta(t = 0^+)$, independent of the way the known state was prepared. This allows the known state to be prepared by an arbitrary input signal so long as the output signal is recorded only after the input is turned off. The two scattering matrices differ only by the identity; the real parts of the scattering-matrix poles appear as dips for Eqn. \ref{eq:scattering_with_input_on} and  peaks for Eqn. \ref{eq:scattering_with_input_off}. The equations presented above capture how coherent states of the electromagnetic field are modified by the dynamics of the lattice and therefore all the experimental results presented here. This approach can be readily extended to arbitrary states by promoting the coherent state amplitudes to operators and including the noise operators injected by the intrinsic decay channels in the scattering relation.

Measured spectra corresponding to Eqns. \ref{eq:scattering_with_input_on} and \ref{eq:scattering_with_input_off} are displayed in Fig. \ref{fig:spectra}. Spectra in Figure \ref{fig:spectra}b  are obtained by continuous driving and flux modulation, and are measured with a VNA over an $8$ MHz bandwidth about site $n_\text{abs} = 32$, corresponding to  $n = 0$. As the modulation frequency is tuned through the average FSR, Floquet quasi-energy peaks associated with nearby sites approach the center of site 32, where couplings can be inferred from avoided crossings. The distortion of mirror symmetry in the plot is due to both on-site frequency disorder and the systematic variation of coupling rates with mode number. This variation is smaller at larger mode numbers but is still appreciable at $n_\text{abs} = 32$. The magnitude of the coupling rates peaks far below site 32 and decreases as $1/n_\text{abs}$ for larger $n_\text{abs}$ (see Appendix \ref{app:coupling_rates}); larger coupling-to-loss ratios favor the appearance of Floquet peaks from lower sites which appear as diagonal streaks across the detuning plot from bottom-left to top-right. Site $32$ was chosen for this measurement because it lies in the center of a $12$-site sublattice that contains no ``barrier sites,'' or sites with severe loss and disorder that act as barriers to the propagation of tight-binding photons.

Fig. \ref{fig:spectra}c contains scattering parameters between a common input site and three neighboring output sites, calculated in the rotating frame. Spectra were measured by first exciting site $n = 0$ with a long pulse from one AWG channel while no modulation is applied, switching on the modulation shortly after the excitation pulse ends, and then detecting the output field from each site (this scheme is described in more detail in Section \ref{sec:transient_prop} and the Appendix). The Fourier transform of the time series at site $j$ is then proportional to the scattering parameter $\Sbold_{j,0}\ofw - \delta_{j,0}$, where the direct-reflection term is automatically subtracted because the incident field is turned off during data collection. 

\section{Site-resolved transient measurements} \label{sec:transient_prop}

\begin{figure}[t]
\includegraphics{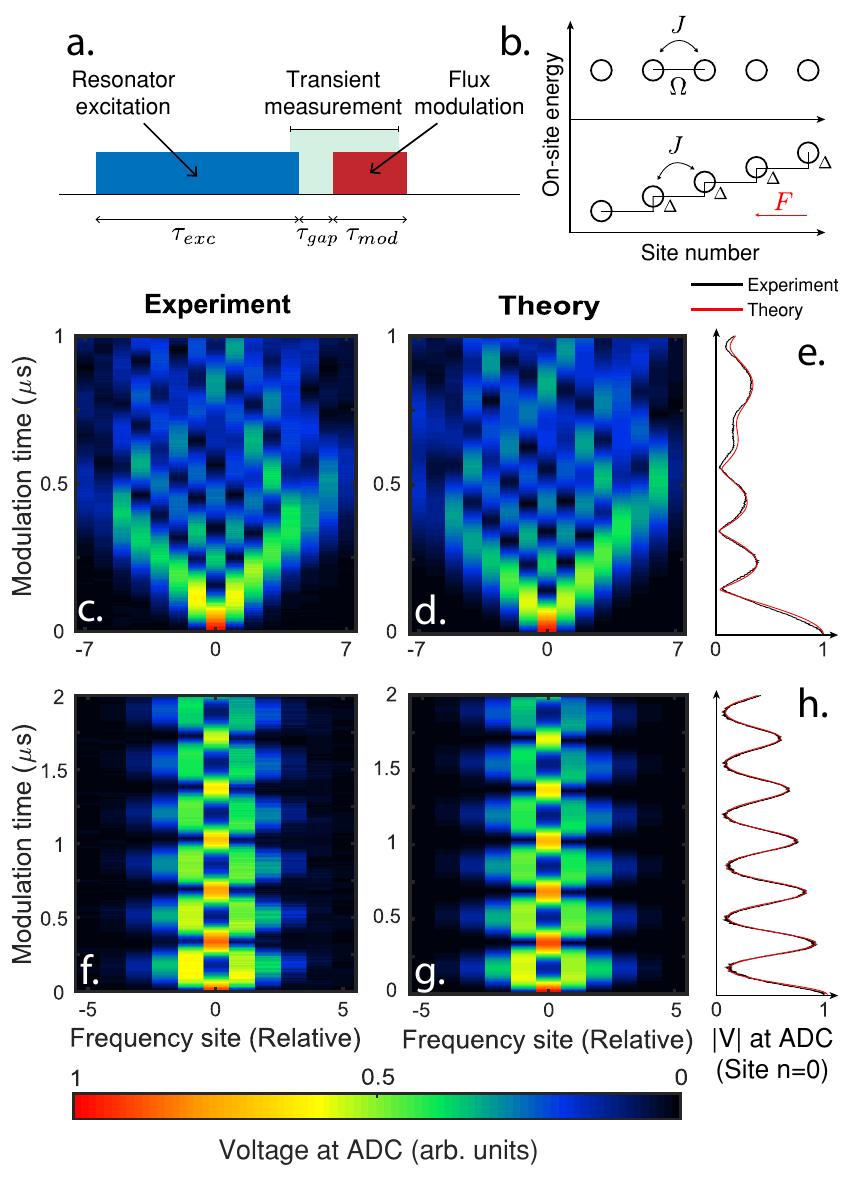}
\caption{\label{fig:single_site_ex} \textbf{Transient propagation from a single-site initial state}. \bpA Pulse schematic for exciting a single mode to steady-state, applying flux modulation and measuring the output voltage while the excitation propagates and rings down. All voltage traces in (\textbf{C} - \textbf{H}) are displayed from the arrival of the modulation pulse to an intermediate time before the end of the modulation pulse and smoothed with a 16-point moving average. The numerical values are the voltage amplitudes $|V_\text{ADC}|$, normalized so that each colormap has a maximum value of $1$. \bpB Schematic of on-site energies in the rotating frame, for modulation resonant with the free spectral range (top) and detuned by an amount $\Delta$ (bottom). We show the ideal case with no on-site disorder.  (\textbf{C} , \textbf{F}) Experimental magnitude of slowly-varying voltage envelopes with flux modulation at (155.1 , 152.1) MHz respectively, compare the theory calculations in (\textbf{D} , \textbf{G}) and the overlaid traces at n=0 in (\textbf{E} , \textbf{H}).}
\end{figure}

\subsection{\label{sec:rot_frame_meas} Rotating-frame measurements}

We observe lattice-coupling dynamics in the time domain by exciting a single initial site and measuring the slowly-varying voltage envelope at an array of adjacent sites. The general scheme of the measurement is shown in Fig. \ref{fig:single_site_ex}a. A long monochromatic pulse ($\tau_\text{exc} = 30~\mus$) drives mode $0$ of the unmodulated lattice to steady-state, exciting site $n=0$ of the lattice. The excitation is switched off and mode $0$ begins to ring down during a short gap $\tau_\text{gap} \sim 150~\text{ns}$. At this point the parametric modulation is turned on, resulting in dynamics for a duration $\tau_\text{mod} \sim 4-32 \, \text{ns}$.  The gap between the excitation and modulation ensures that the modulation arrives at the chip \textit{after} the excitation pulse cuts off. Data collection begins just before the end of the excitation pulse and continues until just before the modulation ends, allowing observation of the steady-state, gap and modulated regimes so $t=0$ when modulation arrives at the chip can be easily identified. 

We measure the output voltage from each frequency site in the rotating frame of eqns. \ref{eq:rot_frame_evol} and \ref{eq:rot_frame_output} using $\an(t) = \bn(t)e^{-i(\wo + n \Wmod)t}$, where $\wo/\tpi \approx 4.9892$ GHz is the center frequency of the uncoupled mode labeled $n = 0$. For each site the corresponding carrier frequency is converted to $125~\text{MHz}$ using a custom downconversion board \cite{Arrangoiz-Arriola2019, Wang2019}, data is collected at the ADC and digitally downconverted to generate a signal at DC. Voltage traces for each site are collected in a raster fashion by shifting the analog downconversion frequency, and the traces are concatenated to form the colormaps in Figs. \ref{fig:single_site_ex}(c,f). All experimental traces shown are smoothed with a 16-point moving average before taking the magnitude, equivalent to multiplication by $\sinc(\pi \omega / (\omega_\text{samp}/16))$ in the Fourier domain.

\subsection{\label{sec:resonant_mod_dynamics} Resonant modulation}
Fig. \ref{fig:single_site_ex}c and d display experiment and theory for propagation driven by resonant modulation at $155.1~\text{MHz}$. The envelope spreads out in a light cone for the first $0.5~\mus$, at a rate bounded by the maximum local group velocity \cite{Jurcevic2014}: $|v_{\text{g}, n}^\text{max}| \approx 2|J_{n}|$ sites per unit time. The light cone tilts slightly to the left due to the systematic variation in coupling rates, which scale as $1/n_\text{abs}$. Near $t = 0.5~\mus$ the envelope reaches two ``barrier sites'' at $n = -6$ and $7$ whose large on-site disorder and loss cause reflections; little signal is emitted beyond the barriers. The theory calculation in Fig. \ref{fig:single_site_ex}d simulates eqn. \ref{eq:time_domain_rot_frame} using $\{\Delm, \kapm \}$ extracted from fitting uncoupled-mode peaks and $\{J_{m, m+k}\}$ predicted from the flux-tuning parameters in Table \ref{tab:tuning_fit_params}. The modulation amplitude in flux quanta is the only free parameter and was chosen as $0.062\Phio$ to best match the trace at $n = 0$ (Fig. \ref{fig:single_site_ex}e). The main uncertainty in the model is due to reflections at the barrier sites, which are not well modeled due to greater uncertainty in fitting the parameters of the barrier modes. Data is displayed up to $t = 1.0~ \mus$ after which reflections from these barriers interfere at the center and the experimental result diverges significantly from theory.

\subsection{\label{sec:detuned_mod_dynamics} Detuned modulation: Bloch oscillations}
We next implement the synthetic-dimension Bloch oscillations proposed in \cite{Yuan2016a}. Figs. \ref{fig:single_site_ex}(f,g,h) display analogous experimental and theory traces for detuned modulation at 152.1 MHz. In the rotating frame this detuning becomes a linear gradient in on-site energy, $\Deln \approx n \Delta$, which simulates a charged particle moving on a lattice in the presence of a spatially-uniform field (Fig. \ref{fig:single_site_ex}b). This field drives the frequency-space quasimomentum $k_f$ according to the Bloch acceleration theorem \cite{Hartmann2004, Grecchi2001, Kittel1987}:
\begin{equation}
    \pdt k_f = F = -\frac{\delta \Deln}{ \delta \wn} \approx -\frac{\Delta}{\Wmod}
\end{equation}
For compactness we define a dimensionless quasimomentum $k \equiv k_{f} \Wmod$ that evolves according to $k(t) - k(0) = -\Delta t$. Due to the periodicity of extended Brillouin zones, an equivalent $k$ is reached after a time $T_{B} = \tpi/|\Delta|$, and lattice dynamics are periodic with frequency $f_{B} = 1/T_{B} = |\Delta|/\tpi$.

We observe symmetric Bloch oscillations about site $n = 0$, with recurrences at the detuning frequency $|\Delta|/\tpi \approx 3~\text{MHz}$. The single-site initial excitation ensures symmetric oscillations as its spatial Fourier transform has uniform amplitude for all $k$ and therefore has no preferred group velocity.
Comparing Fig. \ref{fig:single_site_ex}e to \ref{fig:single_site_ex}h, we see a better agreement between theory and experimental voltage traces in the Bloch oscillation case. This is because the field amplitudes in the Bloch oscillation case are confined to sites $n \in [-3,3]$ where calculation parameters come from more reliable fits.
\section{Dynamics of wavepackets} \label{sec:multi_site}

Single-site excitation leads to the initial condition of a photon localized to a single lattice site without a well-defined momentum. The spatially symmetric character of such a state, as well as the approximate translational and reflection symmetries of the lattice, causes the photon to spread out in both directions. Directional propagation requires the state of the field to be confined in $k$-space so that it samples only a small range of group velocities. It is  possible to initialize the electromagnetic field in a wavepacket with well-defined momentum in a particular direction. In this section, we first measure the dispersion of the lattice, and then outline and demonstrate a scheme for generating and observing propagation of photon wavepackets in the synthetic lattice.

\begin{figure}
\includegraphics{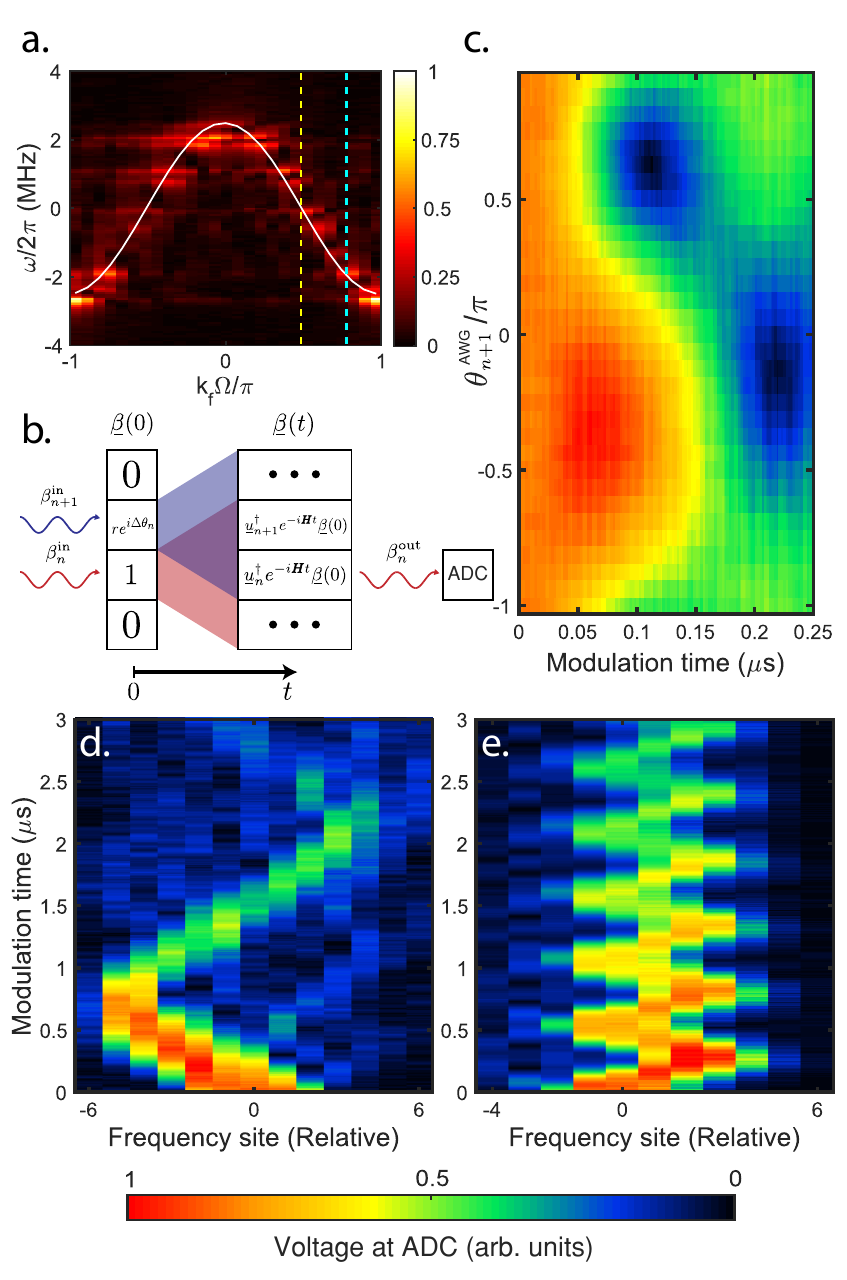}
\caption{\label{fig:multi_site_ex} \textbf{Calibration and propagation of an initial wavepacket}. \bpA Approximate dispersion along the frequency dimension, extracted from peaks in the 2D Fourier transform of transient propagation data as in Fig. \ref{fig:single_site_ex}b. Data is arbitrarily normalized to a maximum of 1. Dashed vertical lines indicate target $k$ values for the initial states in \bpD (yellow) and \bpE (cyan). \bpB Schematic of the nearest-neighbor interference experiment used to calibrate the on-chip amplitudes $\beta(0)$. Interference occurs in the (purple) overlap between light cones emanating from neighboring sites. \bpC Representative calibration data, with $r \approx 1$ for maximum interference amplitude. Sites $-1$ and $0$ are excited by a two-frequency pulse with phase offset $\theta_\text{AWG}$ defined at the start of the pulse. Output voltage amplitude from site $-1$ is plotted for short modulation times as the phase offset is swept. \bpD Propagation of a five-site wavepacket with initial $\keff \approx 0.5 \pi$, showing negative group velocity and coherent reflection of the wavepacket at the n = -6 barrier site. \bpE Directional Bloch oscillation of a 5-site wavepacket with initial $\keff \approx 0.78 \pi$.}
\end{figure}

\subsection{\label{sec:dispersion} Tight-binding dispersion}

In order to excite wavepackets with well-defined propagation characteristics, we need to better understand  the dispersion properties of the synthetic lattice. In an infinite and disorder-free lattice, diagonalizing the Hamiltonian leads to a series of spatially extended and periodic Bloch function eigenstates ($\beta^{(k)}_j(t) = e^{ikj-i\omega(k)t}$) with corresponding wavevectors $k$ and natural frequencies $\omega(k)$. Despite the differences between such a lattice and the experimentally realized one, in particular the finite extent of the experimentally realized lattice as well as the presence of disorder and loss, the same picture approximately holds and a dispersion relation $\omega(k)$ can be extracted. We first initialize the lattice at $t=0$ with an excitation $\betam(0) = \delta_{m,\no}$. Note that this state overlaps with all of the Bloch functions in the infinite perfect lattice case, and that all of the wavevectors play a role in the time-evolution of this state.

We perform a 2D Fourier transform on the full state $\beta_j(t)$ to obtain $\beta[k,\omega]$. This is equivalent to taking the overlap in both time and space of the resulting state with the Bloch functions. For an infinite disorder-free lattice, the result would tend to $\beta[k,\omega] \rightarrow \sum_{k_0} A_{k_0}\delta(k-k_0)\delta(\omega-\omega(k_0))$, i.e. a function that is peaked along the dispersion relation $\omega(k)$ and zero elsewhere. 
Fig. \ref{fig:multi_site_ex}a displays the Fourier transform of a transient propagation measurement with nearest-neighbor coupling, following the schematic of Fig. \ref{fig:single_site_ex}. Traces were collected for sites $n \in [-8, 18]$ with a modulation duration of $15.934 \mus$. The peak Fourier amplitudes cluster around the cosine dispersion of an ideal lattice with nearest-neighbor coupling,
\begin{equation} \label{eq:cosine_disp}
    \omega(k) = 2 |J| \cos(k + \thmod)
\end{equation}
where $\thmod \equiv \arg(J)$.
Comparing the experimental result to the dispersion relation of a perfect lattice, we find that finite length and disorder smears the band diagram in the $k$ dimension, while the finite lifetime of the modes causes blurring in the $\omega$ direction. Fitting the positions of the peak amplitudes to eqn. \ref{eq:cosine_disp} extracts a representative local coupling rate $|J_\text{fit}/\tpi| = 1.25 \text{ MHz}$.

Furthermore, we adjust the measured output field to compensate for output coupling and propagation through the measurement channel:
\begin{equation}
\label{eq:beta_meas_prop}
\betan^\text{meas} = - \sqrt{\kapn^{e}}G^\text{out}(\wn)e^{i\wn \Tout} \betan
\end{equation}
We correct for the output couplings $\kapn^{e}$ using fitted parameters from VNA spectra, recorded with modulation turned off. The output gain $G^\text{out}$ contains the insertion losses and amplifier gains of all components between the chip and ADC as a function of lab-frame frequency $\wn$. We neglect the frequency dependence of $G^\text{out}$ as the lattice disorder appears to be the limiting factor when approximating dispersion in this way. The linear phase shift assumes a uniform propagation delay $\Tout$ and superimposes a phase gradient of $\delta\theta/\delta n = -\Wmod \Tout$ on the chip output phases, effectively shifting the $k$-axis of the Fourier transform so the ideal dispersion is measured as $2|J|\cos(k + \thmod + \Wmod \Tout)$. Fig. \ref{fig:multi_site_ex}a absorbs $\Wmod \Tout$ and $\thmod$ into $k$ because the former is a measurement artifact and the latter can be removed by a unitary transformation \cite{Girvin2014} when there is only one modulation frequency. Crucially, both the excitation and modulation phases must be stable over the duration of the entire experiment or the Fourier transform will be additionally blurred along the $k$-axis. We correct for phase drifts using a reference signal from the AWG, sent to a second ADC channel as described in Appendix \ref{app:exp_details}. In eqn. \ref{eq:beta_meas_prop} we have ignored the dispersion (frequency-dependent group delay) in the measurement channel.

\subsection{\label{sec:multiplex_calib} Calibrating phases for multi-frequency pulses}
To realize initial states with a well-defined momentum, we engineer precise phase relationships between the excitation amplitudes at different sites. This requires us to calibrate the gain and phase response of the input lines. Here, we describe a calibration procedure that involves a sequence of two-site phase measurements and provides the information required to generate driving signals that result in multi-site excitations with a well-defined momentum.
We model the driving signals as superpositions of monochromatic voltage waveforms emitted from the AWG, where each component has the form
\begin{equation}
    v_{n}(t) = 2 \lbar \betan^\text{AWG} \rbar \cos \lp \wn t + \thawg_{n}\rp.
\end{equation}
The driving frequencies $\wn$ are chosen as the fitted center frequencies of the unmodulated modes $n$, such that the resulting field in mode $n$ becomes 
\begin{equation}\label{eq:beta_AWG}
    \betan(t=0) = \sqrt{\kapn^{e}}G^{in}(\wn)e^{-i\theta_n} \left |\betan^\text{AWG} \right|
\end{equation}
when the modes $n$ are spectrally well-resolved. The phase $\theta_n$ equals $\thawg_{n}$ plus an unknown $n$-dependent offset arising from propagation delays and detunings from resonant driving. The detuning phases represent experimental error and are typically small, whereas the propagation phases are much larger than $2\pi$. It is therefore necessary to calibrate the unknown phase offset for each $n$.

To perform this calibration, we excite pairs of adjacent sites with different relative phases, then turn on nearest-neighbor coupling and measure the interference between fields in the two sites as represented schematically in Fig. \ref{fig:multi_site_ex}b. For each pair we set $\thawg_n = 0$, and sweep $\thawg_{n+1}$ from $-\pi$ to $\pi$. The known phase difference at the AWG results in an unknown phase difference $\Delta \theta_n \equiv \theta_n - \theta_{n+1}$ between the on-chip fields, following eqn. \ref{eq:beta_AWG}. We then turn on a resonant modulation signal at $155.1~\text{MHz}$ and measure the voltage amplitude emitted from the lower site $n$. At small modulation times $t \ll 1~\mus$ we observe a sinusoidal interference pattern in the voltage amplitude as $\thawg_{n+1}$ is swept; a representative measurement of this type is shown for $n = -1$ in Fig. \ref{fig:multi_site_ex}c.

We perform the following calculation to understand the interference pattern. For general initial amplitudes $\beta(0)$ and quadratic-form Hamiltonian $\Hbold$ the amplitude at site $n$ evolves as
\begin{equation}
\label{eq:time_evol}
    \betan(t) = \udagn e^{-i \Hbold t} \betaul(0)
\end{equation}
where vectors $\un$ form an orthonormal frequency-site basis and function analogously to Dirac kets $|n\rangle\equiv |\cdots 0_{n-1} 1_{n} 0_{n+1} \cdots \rangle$ in a single-photon manifold. For a two-site interference experiment we let $\betan(0) = 1$ and $\beta_{n+1}(0) = re^{i \Delta \theta_{n}}$ with all other $\betam(0) = 0$; the arbitrary overall scaling does not affect the shape of the interference pattern. The field amplitude in the lower site $n$ can then be expressed as
\begin{multline} \label{eq:neighbor_interf_1}
    |\beta_{n}(t)|^2 = |\Un(t)|^2 + r^2 |\Unp(t)|^2 
    \\
    - 2 \text{Re} \lb i r  \Un(t)^{*} \Unp(t) e^{i(\Delta\theta_{n} + \thmod)}\rb
\end{multline}
where the functions $\mathcal{U}_{n}$ and $\mathcal{U}_{n+1}$ are real-valued when on-site detuning and loss are negligible ($\Deln t \ll 1, \kapn t \ll 1$); see discussion in Appendix \ref{app:neighbor_interf_deriv}. In this limit the interference term becomes $2 r  \Un(t) \Unp(t) \sin(\Delta\theta_{n} + \thmod)$. We absorb the modulation phase into the unknown $n$-dependendent offset to obtain a calibration equation:
\begin{multline}
\label{eq:calib_eqn}
    |\beta_{n}(t)|^2 = |\Un(t)|^2 + r^2 |\Unp(t)|^2 
    \\ 
    + 2 r  \Un(t) \Unp(t) \sin(- \thawg_{n+1} + \theta^{\text{calib}}_{n})
\end{multline}
The calibration therefore amounts to determining the phases $\{\theta^{\text{calib}}_{n}\}$. The measured output field $\betan^\text{meas}(t)$ is proportional to $\betan(t)$ according to eqn. \ref{eq:beta_meas_prop}, so the sinusoidal dependence on $\thawg_{n+1}$ can be directly observed in voltage traces as in Fig. \ref{fig:multi_site_ex}c, which displays $|\betan^\text{meas}(t)|$ during the first $0.25~\mus$ of modulation time for different values of the phase $\thawg_{n+1}$. Fitting the amplitude along a vertical slice at sufficiently small $t$, e.g. $t = 0.1~\mus$, yields $\theta^{\text{calib}}_{n}$ for each pair of sites.

We need $N-1$ such pairwise calibrations to excite $N$ adjacent modes with an arbitrary phase distribution, thus creating a wavepacket with tunable momentum. We are particularly interested in states localized in $k$-space about a target value
\begin{equation}
    \keff \equiv k + \thmod = \frac{\Delta \theta_{n}}{\Delta n} + \thmod.
\end{equation}
The Fourier transform in Fig. \ref{fig:multi_site_ex}a is plotted with respect to this $k_\text{eff}$, and we define the localized $k$ as a discrete phase gradient in analogy to the Bloch wavefunctions $\beta^{(k)}_n(t=0) = e^{ikn}$. But $\Delta n = 1$ for adjacent sites, and we recognize $\Delta \theta_{n} + \thmod = - (\thawg_{n+1} - \thawg_{n}) + \theta^{\text{calib}}_{n}$, which was used to obtain eqn. \ref{eq:neighbor_interf_1} from eqn. \ref{eq:calib_eqn} for the special case where $\thawg_{n} = 0$. By setting this expression constant for all excited sites $n$, we create states localized about 
\begin{equation}
    \keff = - (\thawg_{n+1} - \thawg_{n}) + \theta^{\text{calib}}_{n}.
\end{equation}
Finally, we estimate the value of $r$ by comparing experimental interference patterns with solutions of the time-evolution eqn. \ref{eq:time_evol}. As $r \propto |\beta_{n+1}^{\text{AWG}} / \beta_{n}^{\text{AWG}}|$, determining the proportionality constant allows us to initialize arbitrary amplitude distributions for $N$-site wavepackets.

\subsection{\label{sec:wavepacket_prop} Wavepacket propagation}
We initialize five-site wavepacket states with approximately uniform phase gradients and observe directional propagation for both resonant and detuned modulation. In both cases the target amplitudes for the initial wavepacket form a truncated Gaussian distribution with mean $\mu = 0$:
\begin{equation}
    |\beta_n(t=0)| = \left\{\begin{matrix}
\mathcal{N}_{0}e^{-\frac{n^2}{2\sigma^2}}, & |n| \leq 2
\\ 
0, & |n| > 2
\end{matrix}\right.
\end{equation}

Resonant and detuned propagation are shown in Fig. \ref{fig:multi_site_ex}(d, e) respectively. In Fig.~\ref{fig:multi_site_ex}(d) the target wavepacket has $\sigma = 2.5$ and $k_\text{eff} = 0.5\pi$, with modulation at $\Wmod / \tpi = 155.1 ~ \text{MHz}$ The expected negative group velocity is observed for $t \lesssim 0.7~\mus$ at which point the wavepacket reflects off the barrier site at $n = -6$ and the group velocity becomes positive. Predictions for reflection and transmission coefficients at a barrier site are discussed in Appendix \ref{app:elastic_scattering_freq}. The modulation amplitude was reduced by half ($0.031 \Phio$) to increase the number of samples taken before the wavepacket reflects; dividing the value of $J_\text{fit}$ from section \ref{sec:dispersion} by 2 predicts a group velocity of $-7.85~\text{sites/}\mus$. At this group velocity the center of a wavepacket would take $0.6~\mus$ to reach the lowest pre-barrier site ($n = -5$), consistent with Fig.~\ref{fig:multi_site_ex}(d).

In (e) the target wavepacket has $\sigma = 2$ and $\keff = 0.78\pi$, with modulation $\Wmod / \tpi = 153.1$ MHz. This initial $\keff$ was chosen such that the wavepacket avoids barrier sites throughout a Bloch period. Slightly less than 6 Bloch periods are observed in $3~\mus$, indicating an effective detuning slightly smaller than $2$ MHz. The maximum modulation amplitude of $0.062 \Phio$ was used, so the maximum group velocity during a Bloch period is twice that of (d) at roughly $15.7~ \text{sites/}\mus$. In both (d) and (e) a small amount of bidirectional propagation is observed starting immediately after $t=0$, suggesting the initial wavepackets have components in $k$-space far from the target $\keff$. We attribute these parasitic components to the limited spatial extent of the wavepacket; we have truncated the Gaussian envelope to zero outside the initial 5 sites.
Bidirectional propagation can be reduced by initializing a wavepacket with larger spatial extent and smoother decay to zero.
\section{Discussion} \label{sec:discussion}
We have demonstrated tight-binding coupling along a synthetic dimension in the upper modes of a superconducting circuit and observed the propagation of classical site amplitudes both bidirectionally and with a strongly-directional group velocity. In addition to the low losses and large hopping rates demonstrated, the advantage of this platform is its compatibility with strongly nonlinear superconducting qubits, and tailorable parametric interactions, both of which provide an avenue to implementing hardware-efficient analog quantum simulation. For our multi-mode resonator, dynamics in the synthetic dimension are driven through a single flux-control channel and observed through a single external-coupling channel which simplifies the hardware requirements of packaging and wiring the device in a dilution refrigerator. The main performance-limiting factors in this work are the typical loss rates at each site ($\kapn/\tpi \sim 100 \text{ kHz}$) and the presence of high-loss, high-disorder ``barrier sites'' that confine well-controlled dynamics to a smaller number of sites. Improved design and fabrication methods are the primary avenues for reducing disorder and loss.

In addition to the nearest-neighbor coupling presented in this work, a rich array of additional experiments are immediately accessible in this platform. Customizable band structure and synthetic gauge fields can be implemented by multiplexing the flux modulation with signals near integer multiples of the free spectral range, as observed experimentally in \cite{Dutt2019}. The coupling dynamics of the Hamiltonian can be effectively time-reversed by a sudden shift of $\pi$ in the modulation phase \cite{Minkov2018} allowing revival of an initial state (limited by on-site disorder and loss rates); a preliminary experimental investigation is reported in the Appendix. Degenerate or non-degenerate parametric oscillation can be implemented in any mode(s) $\an$ and $\am$ by modulating at frequencies near $\wn + \wm$ \cite{Wustmann2013, Wustmann2017, Wang2019}, where the degenerate case corresponds to $m = n$. Such a multi-mode parametric oscillator could be utilized to generate continuous-variable cluster states as proposed in \cite{Menicucci2008, Pfister2019}, which provide a starting point for one-way quantum computation. The resonator may also be operated in the nonlinear regime by increasing the mean intracavity photon number and/or fabricating a shorter SQUID array.  Further applications arise from coupling multi-mode resonators to each other to provide a spatial dimension, or coupling a qubit to a multi-mode resonator at one frequency site to simulate an electronic system coupled to a bath with tunable density of states.

\begin{acknowledgments} \label{sec:acknowledgements}
The authors would like to thank C. Sarabalis, R. Patel, J. Witmer, R. Van Laer, A. Dutt, C. Wojcik, and S. Fan for useful discussions. This work was supported by a MURI grant from the U. S. Air Force Office of Scientific Research (Grant No. FA9550-17-1-0002). N.R.A.L. was partially supported by a Stanford Graduate Fellowship. Part of this work was performed at the Stanford Nano Shared Facilities (SNSF), supported by the National Science Foundation under Grant No. ECCS-1542152, and the Stanford Nanofabrication Facility (SNF).
\end{acknowledgments}

\appendix

\section{Device fabrication} \label{app:fab}
The 14mm$\times$14mm device was fabricated with a 6-mask lithography process on a 500-$\mu m$ high-resistivity Si substrate ($\rho > 20 k\Omega \cdot cm$). Mask 1 defines the aluminum ground planes, CPW resonator and feedlines (all 150nm thick), patterned by image-reversal photolithography and liftoff. Mask 2 adds palladium alignment marks by the same image-reversal method, in preparation for aligning the electron-beam lithography mask that will later create the SQUID array. Masks 3 and 4 create superconducting aluminum airbridges (300nm thick) using the procedure of Ref. \cite{Chen2014}. Mask 3 defines bridge feet and spans using reflowed thick photoresist, 300 nm Al is evaporated, and mask 4 protects the metal bridge spans while a wet etch removes the rest of the aluminum. Mask 5 patterns the SQUID array using a Dolan-bridge, double-angle technique for growing Al/AlO$_x$/Al junctions via \textit{in situ} oxidation \cite{Dolan1977, Kelly15}; this technique is only use of electron-beam lithography in the process. Mask 6 uses a bandage process to form superconducting aluminum connections between the SQUID array, CPW resonator and ground plane \cite{Dunsworth2017}. Test arrays with SQUID numbers between 0 and 10 are fabricated elsewhere on the chip in Mask 5 and connected to bond pads from Mask 1 by the bandage process. This allows measurement of normal-state resistances (which extracts the approximate Josephson inductance via the Ambegaokar-Baratoff equation \cite{Ambegaokar1963}), and scanning electron microscopy of the arrays without damaging the experimental device.

Fabricating the 3-$\mu m$ high airbridges before the SQUID arrays leads to systematic distortion of the bridge spans. The electron-beam resist forms a bilayer approximately $1 \mu m$ thick, which spreads partway up the bridge spans but does not cover the maximum height. The sides of the bridge spans crimp inward during the subsequent lithography steps, perhaps due to strain in the resist that affects only the lower portion of the spans. The center of each span is parasitically exposed to aluminum evaporation from the double-angle process, which is thin enough (30nm + 50nm) that it tears away from the bridge spans during liftoff without destroying the span. However, both crimping and parasitic deposition add disorder to the bridge shapes and create small crevasses where resist can become trapped, increasing both on-site disorder and loss in the device Hamiltonian. This distortion can be minimized in future devices by rearranging the fabrication process such that airbridges are patterned last, as their shapes typically survive the thick photoresist used in the final dicing process.

\section{Experimental details} \label{app:exp_details}

\begin{figure*}
\includegraphics{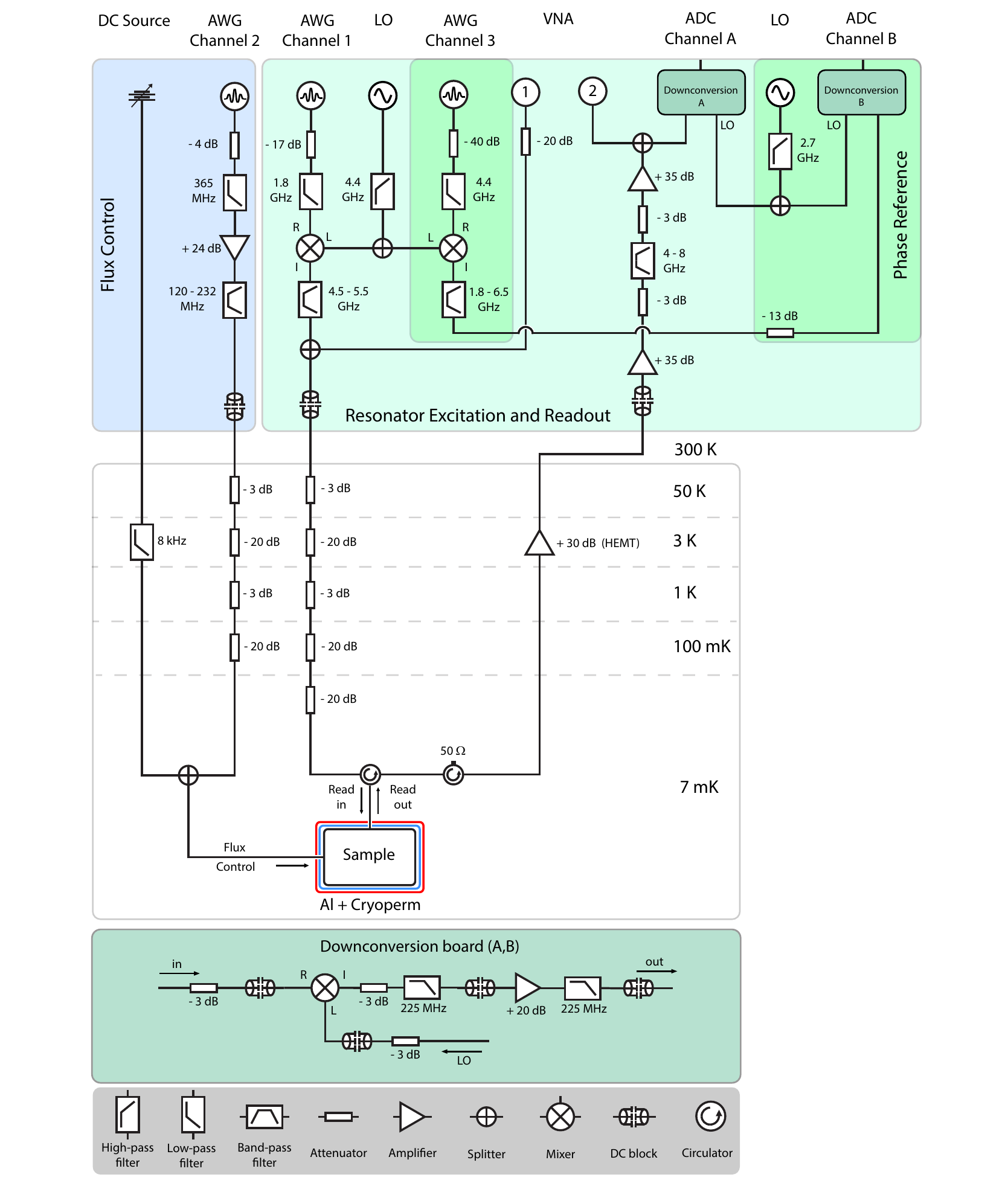}
\caption{\label{fig:exp_setup} \textbf{Experimental setup.} The sample is located at the mixing-chamber plate of a dilution refrigerator, packaged in a
microwave PCB and copper enclosure, and surrounded by cryogenic magnetic shielding. All instruments are phase-locked by a
10 MHz rubidium frequency standard (SRS SIM940); the output of both local oscillator sources is high-pass filtered to remove feedthrough from the 10 MHz clock. Notation of this figure follows Refs. \cite{Arrangoiz-Arriola2019, Wang2019}.}
\end{figure*}

\subsection{Resonator Excitation} \label{appsub:exp_overview}
The experimental setup is shown in Fig. \ref{fig:exp_setup}. For all pulsed experiments in this work, we generate resonator excitation and flux modulation using a 5 GS/s arbitrary waveform generator (AWG) (Tektronix series 5200). The target resonances of the device lie between 4.5 and 5.5 GHz, which we produce using difference-frequency upconversion with a dedicated local oscillator (Keysight E8257D) and triple-balanced mixer (Marki T3-20). Long square-pulse envelopes (typically 30$\mus$) are used to drive resonances to steady-state before the modulation signal arrives. We operate the AWG in real-waveform mode, in which waveforms are directly synthesized by sampling at 5 GS/s and the built-in digital IQ mixers are not used. While the digital IQ mixers can be used to output frequencies above the 2.5 GHz Nyquist limit, we found empirically that the numerical local oscillators (NCOs) of two different channels acquire a relative time delay when the output frequency of one channel is reset, creating a new and unpredictable phase relationship between the channel outputs which negates the effect of the phase-reference channel discussed in section \ref{appsub:exp_phaseref}. Directly-synthesized waveforms do not experience this problem, but exhibit strong Nyquist images that become difficult to separate from the desired signal if the target frequency is too close to 2.5 GHz. The AWG output is low-pass filtered before the mixer to remove Nyquist images and clock feedthrough, and band-pass filtered after the mixer to remove spurious intermodulation signals and feedthrough from the upconversion LO. In addition to the AWG, a vector network analyzer (VNA) (Rohde $\&$ Schwarz ZNB20) is connected in parallel using 50$\Omega$ resistive power splitters (Mini-Circuits ZFRSC-123+), with insertion losses of approximately 10 dB between the common port and either split port. The VNA is used to collect continuous-wave modulation spectra, and to calibrate the frequencies and widths of uncoupled resonances.

\subsection{Resonator Readout} \label{appsub:exp_readout}
The resonator is driven and read out through a single capacitively-coupled port with $C_{\kappa} \approx 15$ fF and corresponding coupling rates $\kapn^{e}/\tpi$ increasing from approximately 30 kHz to 120 kHz across the experimentally-accessible band of 4 to 8 GHz. The output signal is separated from the input using a three-port circulator at 7 mK (Quinstar QCY-060400C000) and passes through a second circulator functioning as an isolator; the 4-8 GHz circulator bandwidth defines the observable frequency sites. The signal is amplified at 3 mK with a high electron mobility transistor (HEMT) amplifier (Caltech CITCRYO1-12A) and at room temperature by two low-noise amplifiers (Miteq AFS4-02001800-24-10P-4 and AFS4-00100800-14-10P-4). Next, the signal is down-converted to an intermediate frequency (IF) of 125 MHz using a separate local oscillator (Keysight E8257D) and a double-balanced mixer (Marki ML1-0220I). Finally, the IF signal is amplified, low-pass filtered, and digitized by channel A of an acquisition card (AlazarTech ATS9350) with 12-bit resolution and a 500 MS/s sampling rate. The data is first stored on-board and then transferred to a GPU for real-time processing. 

The key readout technique used in this work is a raster scan of the downconversion frequency across equally-spaced frequency sites separated by the fundamental modulation frequency $\Wmod$. To measure the slowly-varying envelope of the output signal at site $n$, the site is assigned a center frequency $\wn' = \wo + n \Wmod$, and the Keysight local oscillator is adjusted to down-convert $\wn'/\tpi$ to 125 MHz. After data is transferred to the GPU, it is digitally down-converted to place one of $\pm$125 MHz at DC, and a smoothing filter is applied after all real-time processing to isolate a single-site envelope within a bandwidth of $< 10$ MHz. Experiments are repeated with the same excitation pulses as the readout site $n$ is varied, providing single-site readout resolution.

\subsection{Flux Control} \label{appsub:exp_flux}
The flux threading the SQUID array is controlled by an on-chip flux line, for which DC and RF currents are combined in a bias tee at 7 mK (Anritsu K250). DC biasing is performed using a programmable voltage source (SRS SIM928), which is low-pass filtered at the 3 K stage (Aivon Therma-24G). RF modulation pulses are generated by the AWG and amplified at room temperature (Mini-Circuits ZX60-P103LN+); attenuation before the amplifier is chosen such that the input power is just below the 1 dB compression point. The signal is low-pass filtered before the amplifier to remove clock feedthrough and Nyquist images, and band-pass filtered after the amplifier to remove harmonics from nonlinear amplification and reduce flux noise.

\subsection{Phase-Reference Channel} \label{appsub:exp_phaseref}
In order to estimate lattice dispersion using a Fourier transform across frequency sites as in section \ref{sec:multi_site}, the data collected at each site $n$ in the downconversion raster scan must be representative of the same experiment and therefore have a consistent phase reference for the carrier frequency. However this phase is not typically stable between measurements. The phase of the downconversion LO is effectively randomized every time the frequency is changed, both LO's experience a gradual phase drift even when nominally locked to an external clock, and we anticipate some jitter in the timing of AWG playback and ADC data collection from experiment to experiment. We summarize the carrier phase of the signal emitted from site $n$ and collected at ADC Channel A as:
\begin{multline} \label{eq:phase_jitter_model}
    \theta_{A}^{n}(t) = -\w_{Ch1} T_{Ch1}(t) + \w_{UC} T_{UC}(t) + \theta_\text{chip}^{n}(t) + 
    \\
    n\Wmod T_{Ch2}(t) - \w_{DC} T_{DC}(t) + \theta_{DC}^{\text{rand}} +
    \\
    \w_{ADC}T_{ADC}(t) + \text{const.}
\end{multline}

Here $T_{X}(t)$ is a reference time for signal generator or detector $X$, and we have written only the terms which we expect to jump or drift between measurements at different sites. Stable phases are lumped into the constant term, including the $n$-dependent delay from linear cable dispersion. We need to provide a reference signal to ADC Channel B that contains the drifting phases such that only the on-chip dynamics remain:
\begin{equation} \label{eq:phase_corrected_ideal}
    \theta^{n} (t) \equiv \theta_{A}^{n}(t) - \theta_{B}^{n}(t) = \theta_\text{chip}^{n}(t) + \text{const.}
\end{equation}
 To accomplish this we note that $T_{ChX}(t) = T_{AWG}(t)$ for all AWG channels when no digital IQ upconversion is used, $\left \{ T_{UC}(t), T_{DC}(t), \theta_{DC}^{\text{rand}} \right \}$ can be replicated up to a constant by splitting the respective LO's between two different mixers, and $T_{ADC}(t)$ is the same for both ADC channels (capture starts simultaneously). The phase of the signal collected by the ``phase reference'' portion of Fig. \ref{fig:exp_setup}, and the resulting ``corrected'' carrier phase is then:
 \begin{multline} \label{eq:phase_refchan_model}
     \theta_{B}^{n}(t) =  -\w_{Ch3} T_{AWG}(t) + \w_{UC} T_{UC}(t) - \w_{DC} T_{DC}(t) 
     \\
     + \theta_{DC}^{\text{rand}} + \w_{ADC}T_{ADC}(t) + \text{const.}
 \end{multline}
\begin{multline}
    \theta_{A}^{n}(t) - \theta_{B}^{n}(t) = (\w_{Ch3} -\w_{Ch1} + n\Wmod ) T_{AWG}(t) 
    \\
    + \theta_\text{chip}^{n}(t) + \text{const.}
\end{multline}
This corrected phase is then stable for measurements at different sites if when measuring site $n$, the AWG Channel 3 output frequency is set to $\w_{Ch1}  - n\Wmod$. We set the reference channel to continuous-wave operation. After each pair of voltage traces is collected, an overall phase $\theta_{B}$ is calculated from the time average of $v_{B}(t)$ and a corrected trace for A is calculated as $v_{A}(t)e^{-i\theta_{B}}$. We emphasize that this correction is made only to facilitate the Fourier transform over frequency sites and does not affect the physics being measured, as any distribution of overall phases $\theta_{n}$ can be removed by a unitary transformation of type $U_{nm} = e^{-i \theta_{n}} \delta_{nm}$ (in the quantum case the exponential would include the operator $\bdagn \bn$). 

A primary caveat is that the reference signal can leak into the resonator excitation line if the RF/LO isolation of the mixers and the isolation between the LO splitter ports are not sufficiently high.
The leaked signal can become upconverted in the process and then coincides with frequency site $n$, meaning it is likely to fall within the 4.5-5.5 GHz passband and parasitically excite the device. We observed such parasitic excitations when the full Channel 3 output amplitude was used, motivating the use of 40 dB attenuation before upconversion to reduce leakage magnitude. The low (6dB) isolation of the upconversion splitter (Mini-Circuits ZFRSC-183+) contributes to the leakage, though we continued to use the splitter because its low insertion loss enables the LO to provide $\sim 17$ dB of power to both mixers. Improved implementation can be achieved using a non-resistive power splitter for better isolation, a LO source with greater output power, and/or mixers requiring lower LO power.

\section{A recent similar work in band structure spectroscopy} \label{app:band_structure_spectroscopy}
A robust procedure for measuring band structure along a synthetic dimension was recently demonstrated by Dutt et. al. in a fiber ring resonator with electro-optic phase modulation \cite{Dutt2019}. The procedure derives the identification of $\kf$ with a time coordinate within a single modulation period, and extracts $\kf(\omega)$ from the timing of transmission peaks when the ring is driven at frequency $\omega$. In principle this strategy is immediately adaptable to superconducting circuits, however it requires a large number of data samples per modulation period, or equivalently $\omega_\text{samp}/\Wmod \gg 1$. This ratio exceeds 300 for the optical experiment but is only ($\frac{500 ~\text{MHz}}{155 ~\text{MHz}} \approx 3.2$) for our device, too small to measure band structure directly from peak timing. We increase the resolution of $\kf$ above 3 samples per Brillouin zone by recording traces at 27 discrete frequency sites, which samples the output spectrum only in the intervals where signal is expected. The ``spatial'' Fourier transform over these sites then approximates a time-domain trace of the output field filtered by the band containing the 27 sites, approximately 3.7 - 8.1 GHz. The measurement shift $\kf \Wmod \rightarrow (\kf + \Tout)\Wmod$ due to output delay is also consistent with the interpretation of $\kf$ as a time.

\section{Rotating frame} \label{app:rot_frame}
We explicitly define the quantities in section \ref{sec:matrix_elements} in terms of circuit parameters and complete the derivation of coupling matrix elements. The modulated coefficient $\bigD(t)$ is given by the effective Josephson energy of the SQUID array at normalized flux $f(t)$, minus its value at the DC flux bias $F$:
\begin{eqnarray}
    H(t) = \sum_n \wn \adagn \an + \bigD(t) \phisqxd
    \\
    \bigD(t) \equiv \frac{\EJ(f(t)) - \EJ(F)}{2}
    \\ \label{eq:EJdef}
    \EJ(f) = \EJo \sqrt{\cos^2\lp f \rp + \dsq^2 \sin^2\lp f \rp}
    \\
    f(t) \equiv F + \df(t) = \frac{\pi}{\Phio}  \lp \Phi_\text{DC} + \Phi_\text{AC}(t) \rp 
\end{eqnarray}
where $f(t)$ is proportional to the flux threading each SQUID loop including DC bias $F$, $\EJ$ is the effective Josephson energy of the SQUID array at flux bias $f$, and each $\wn = \wn(F)$ is fixed to its value at the DC bias $F$. Parameter $\dsq$ describes the asymmetry between junctions on either side of the SQUID array; conditions for the validity of eqn. \ref{eq:EJdef} for $\Nsq > 1$ are discussed in Appendix \ref{app:sq_array}. Collecting the uncoupled oscillator terms as $H_{0}$, expanding $\phixd$ in the uncoupled normal-mode basis and $\bigD(t)$ as a Fourier cosine series:
\begin{multline} \label{eq:pre_rot_H}
        H(t) = H_{0} + \sum_{kmn} \bigD_{k} \cos(k\Wmod t + \thk)  
        \\
        \cdot \phizm \phizn  \lp\am + \adagm \rp \lp \an + \adagn \rp 
\end{multline}
Writing Heisenberg-Langevin equations for weak input-output coupling:
\begin{equation}
    \amdot = -i \lb \am, H \rb- \frac{\kapm}{2}\am  + \sqrt{\kapm^{e}} \am^\text{in}
\end{equation}
and evaluating the commutators:
\begin{multline} \label{eq:pre_rot_heislang}
    \amdot = \lp -i\wm - \frac{\kapm}{2} \rp \am 
    \\
    - 2i \phizm \lp \sum_{kn} \bigD_{k} \cos(k\Wmod t + \thk)  \phizn \lp \an + \adagn \rp \rp
    \\
    + \sqrt{\kapm^{e}} \am^\text{in}
\end{multline}
Authors often use different conventions for the explicit phase in front of $\am^\text{in}$, see for example the seminal development in \cite{Collett1984} and later use in \cite{Girvin2014}.

 We next define a rotating frame by $\an(t) = \bn(t) e^{-i\wn' t}$ where $\wn' \equiv \wo + n\Wmod$. Here is it crucial to distinguish between the equally-spaced rotation frequencies $\{ \wn' \}$ and the uncoupled mode frequencies $\{\wn \}$ which are not in general equally-spaced; disorder is incorporated in the on-site energies $\Deln \equiv \wn - \wn'$. If the input spectrum is similarly localized we can write $\an^{in}(t) = \bn^{in}(t) e^{-i\wn^{in} t}$ (this anticipates classical treatment in which the quantum noise spectrum is neglected). Differentiating the envelope equation, substituting into eqn. \ref{eq:pre_rot_heislang}:
 \begin{equation}
     \andot = \lp \bndot - i\wn' \rp e^{-i\wn' t}
 \end{equation}
 \begin{widetext}
 \begin{multline}
     \lp \bmdot - i\wm' \rp e^{-i\wm' t} = \lp -i\wm - \frac{\kapm}{2} \rp \bmnew e^{-i\wm' t} 
     \\
     - 2i\phizm \lp \sum_{kn} \bigD_{k} \cos(k\Wmod t + \thk) \phizn \lp \bn e^{-i\wn' t} + \bdagn e^{i\wn' t} \rp \rp
     + \sqrt{\kapm^{e}} \bmnew^\text{in} e^{-i\wm^{in} t}
 \end{multline}
 \begin{multline} \label{eq:rot_frame_full_time_dep}
    \bmdot =  \lp -i\Delm - \frac{\kapm}{2} \rp \bmnew - i\phizm \sum_{kn} \bigD_{k}\phizn  \bn \lp e^{-i((n-m+k)\Wmod + \thk)t} + e^{-i((n-m-k)\Wmod - \thk)t}  \rp  ... 
    \\
   + \bdagn \lp e^{-i(((-n-m+k)\Wmod - 2\wo) t + \thk)} + e^{-i(((-n-m-k)\Wmod - 2\wo) t - \thk)}  \rp  + \sqrt{\kapm^{e}} \bmnew^\text{in} e^{-i(\wm^{in} - \wm') t} 
 \end{multline}
 \end{widetext}
 Eqn. \ref{eq:rot_frame_full_time_dep} simplifies greatly under the rotating wave approximation (RWA) which discards all time-dependent terms at frequencies much greater than the other characteristic rates in the system. Here by design we have $\Wmod \gg (|\Delm|, \kapm, |\bigD_{k} \phizm \phizn|)$, so the RWA can be applied to the $\bn$ terms as $e^{-i((n+m+k)\Wmod t} \rightarrow \delta_{n, -m-k}$. The $\bdagn$ terms correspond to one- and two-mode squeezing and are negligible when $k \Wmod \ll (\wm' + \wn')$. In this work we always have $\bigD_{k > 2} \approx 0$ and $\Wmod \ll \wm'$ for all experimentally-accessed sites $m$, so we ignore the squeezing terms. However it is straightforward to drive squeezing with higher modulation frequencies (as in \cite{Wang2019}), and there exist values of $\Wmod/\tpi < 1 \text{ GHz}$ which \textit{are} resonant with squeezing in the lowest modes of the device and must be accounted for or avoided in an optimized experiment. Finally we can set $\wm^{in} = \wm'$ in the input signal to generate the experimental model:
 \begin{multline}
     \bmdot \approx \lp -i\Delm - \frac{\kapm}{2} \rp \bmnew  ...
     \\
     - i \phizm \sum_{k \geq 0}\bigD_{k}  \lp \phi^\text{zp}_{m-k}e^{-i\thk} \hat{b}_{m-k} + \phi^\text{zp}_{m+k}e^{-\thk} \hat{b}_{m+k} \rp  + ...
     \\
     \sqrt{\kapm^{e}} \bmnew^{in}
 \end{multline}
 which is equivalent to eqn. \ref{eq:time_domain_rot_frame} if we redefine the index $k$ over all integers and write the coupling rates as
 \begin{equation} \label{eq:coupling_rates_general}
    J_{m, m+k} = \bigD_{k} \phizm \phi^\text{zp}_{m+k} e^{i \theta_{|k|}  \sgn(k)}.
 \end{equation}
 For $k=0$ we obtain a shift in the on-site frequencies: $\Delm \rightarrow \Delm + \bigD_{0} \phizmsq$, where $\bigD_{0} = \left \langle \bigD(t) \right \rangle_{t}$ may be nonzero due to the curvature of $\EJ(f)$ and is negative for a symmetric SQUID array. Given the flux-tuning parameters in section \ref{sec:flux_tunability} we estimate a nearly-uniform redshift of 200 kHz for the sites accessed in this work.

\section{Zero-point amplitudes} \label{app:coupling_rates}

\begin{figure}[b]
\includegraphics{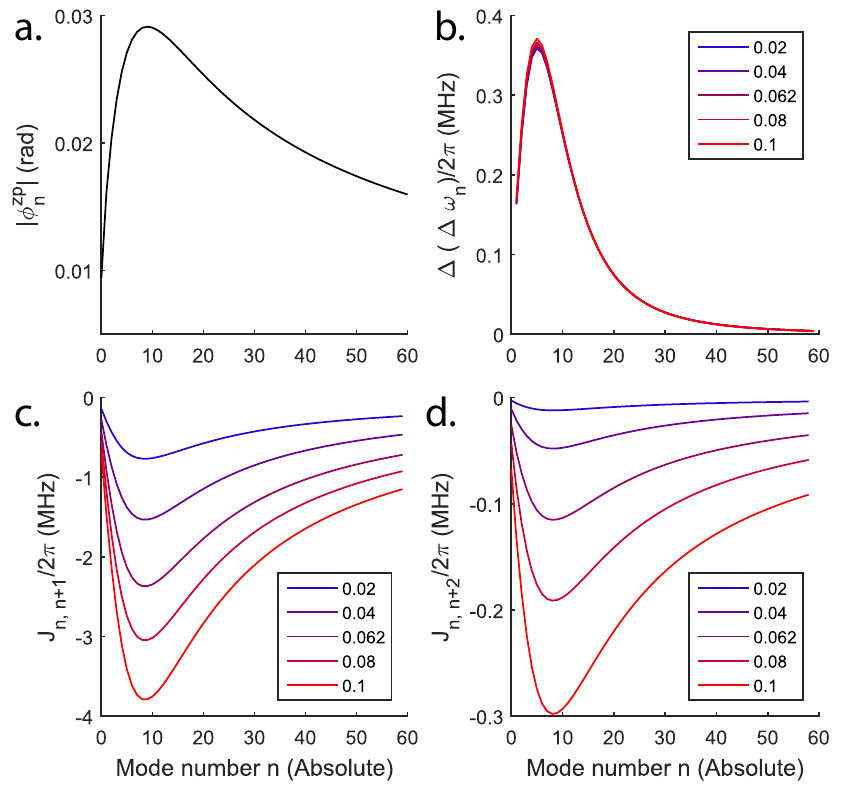}
\caption{\label{fig:phi_zp} \textbf{Zero-point phase distribution and related quantities.} \bpA Predicted phase magnitudes for the first $60$ modes of the device using eqn. \ref{eq:phi_zp_expr}, with a maximum at $n = 9$ followed by decay at higher n. All modes satisfy $|\phizn| \ll 1$. \bpB Variation in free spectral range (FSR) in the absence of fabrication disorder, plotted as a discrete second derivative of the mode frequencies. The FSR increases monotonically with n, stabilizing asymptotically for $n \gg 30$. \bpC Predicted single-hopping rates for single-tone modulation at 155.1 MHz, increasing linearly with modulation amplitude. \bpD Predicted double-hopping rates for single-tone modulation at $155.1 ~\text{MHz}$, increasing quadratically with modulation amplitude. At the experimental value $\Phi = 0.062 \Phio$, $|J_{n, n+2}| < 0.05 |J_{n, n+1}|$, however the double-hopping rates are still used for theory comparisons as in Fig. \ref{fig:single_site_ex} of the main text.}
\end{figure}

Eqn. \ref{eq:coupling_rates_general} indicates that the zero-point phase amplitudes $\phizm$ determine the distribution of coupling rates along the synthetic lattice, with overall scaling and phase given by the modulation signal. Zero-point amplitudes are proportional to the square root of the characteristic impedance of each resonance \cite{Girvin2014}:
\begin{equation}
    \phizm = \frac{1}{\rphio} \sqrt{\frac{\hbar \Zm}{2}} \cos(\ym)
\end{equation}
where $\Zm \equiv \sqrt{L_{m}/C_{m}}$ and $\ym = \km d$. Following \cite{Wallquist2006} we obtain:
\begin{gather}
    C_{m} = \frac{cd}{2} \lp 1 + \frac{\sin(2 \ym)}{2 \ym} \rp + C_{s} \cos^2(\ym)
    \\
    L_{m}^{-1} = \frac{\ym^2}{2ld} \lp 1- \frac{\sin(2 \ym)}{2 \ym} \rp + \frac{\EJ(F)}{\rphio^2} \cos^2(\ym).
\end{gather}
Here $\EJ(F)$ is evaluated at the DC flux bias. Using eqn. \ref{eq:tuning_model} and with some algebra, $\phizm$ can be expressed as
\begin{equation} \label{eq:phi_zp_expr}
    \phizm =  \frac{\sqrt{\hbar \Zo} \cos(\ym)}{\rphio \sqrt{\ym + \lp \frac{\tilde{A}(F)}{\ym} + \frac{\ym}{B}\rp \cos(\ym)}},
\end{equation}
where $\tilde{A}(F) \equiv A \sqrt{\cos^2(F) + \dsq^2 \sin^2(F)}$.
Eqn. \ref{eq:phi_zp_expr} determines the relative size of all parametric modulation terms in the Hamiltonian and offers some insight for device design. $\Zo \equiv \sqrt{l/c}$ is the wave impedance of the CPW; while typically set to $50~\Omega$ to minimize reflections at the input coupler, $\Zo$ can in principle be increased for larger modulation amplitude. All factors in $\phizm$ are positive except $\cos(\ym)$, which introduces a sign parity since $\ym = \frac{\pi \wm}{\wRT}$ increases in steps of approximately $\pi$ and therefore $\cos(y_{m+1}) \approx - \cos(\ym)$. This parity is irrelevant for nearest-neighbor coupling because $\phizm \phi^{zp}_{m+1}$ is always negative, but becomes important when driving more than one type of coupling (e.g. when choosing the relative phase between second- and first-nearest neighbor coupling when both are implemented at once).

The dependence of $\phizm$ on $m$ sets a theoretical limit on the translational invariance of the Hamiltonian. Eqn. \ref{eq:phi_zp_expr} has competing terms with non-periodic dependence on $\ym$, so considering the limiting behavior at small and large mode numbers:
\begin{equation} \label{eq:phi_zp_limits}
\left\{\begin{matrix}
 \phizm \propto \ym^{1/2} & \ym \rightarrow 0 \\ 
 \phizm \propto \ym^{-1/2} & \ym \rightarrow \infty 
\end{matrix}\right.
\end{equation}

These limits follow from the impedance of the SQUID array's LC circuit model. The impedance is primarily (inductive, capacitive) at (low, high) frequencies and approaches zero in either case, shorting the CPW to the ground plane such that $\phizm$ across the SQUID array must also approach zero. The switch in limiting behavior means $|\phizm|$ has a maximum which from eqn. \ref{eq:phi_zp_expr} occurs near $\ym \approx \tilde{A}(F)$. The device measured in this work has $A \approx 40$ and $F \approx -\pi/4$, which predicts a maximum zero-point phase near $\left \lfloor \ym(\text{max}) / \pi  \right \rfloor = 9$. The nearest-neighbor coupling rate $J_{m, m+1}$ also has a maximum near $m = 9$ which explains the appearance of additional Floquet peaks from lower sites (and not upper sites) in Fig. \ref{fig:spectra}b and the slight leftward tilt of the light cones in Fig. \ref{fig:single_site_ex}. Strictly this means the Hamiltonian depends systematically on $m$ and approaches translational invariance only near the maximum and asymptotically as $m \rightarrow \infty$, where coupling rates decay as $1/m$. This decay can be partially compensated by increasing the flux modulation amplitude, constrained by $\Phi_\text{AC} < 0.25 \Phio$.

\section{Model for low-disorder SQUID arrays} \label{app:sq_array}

\begin{figure}
\includegraphics{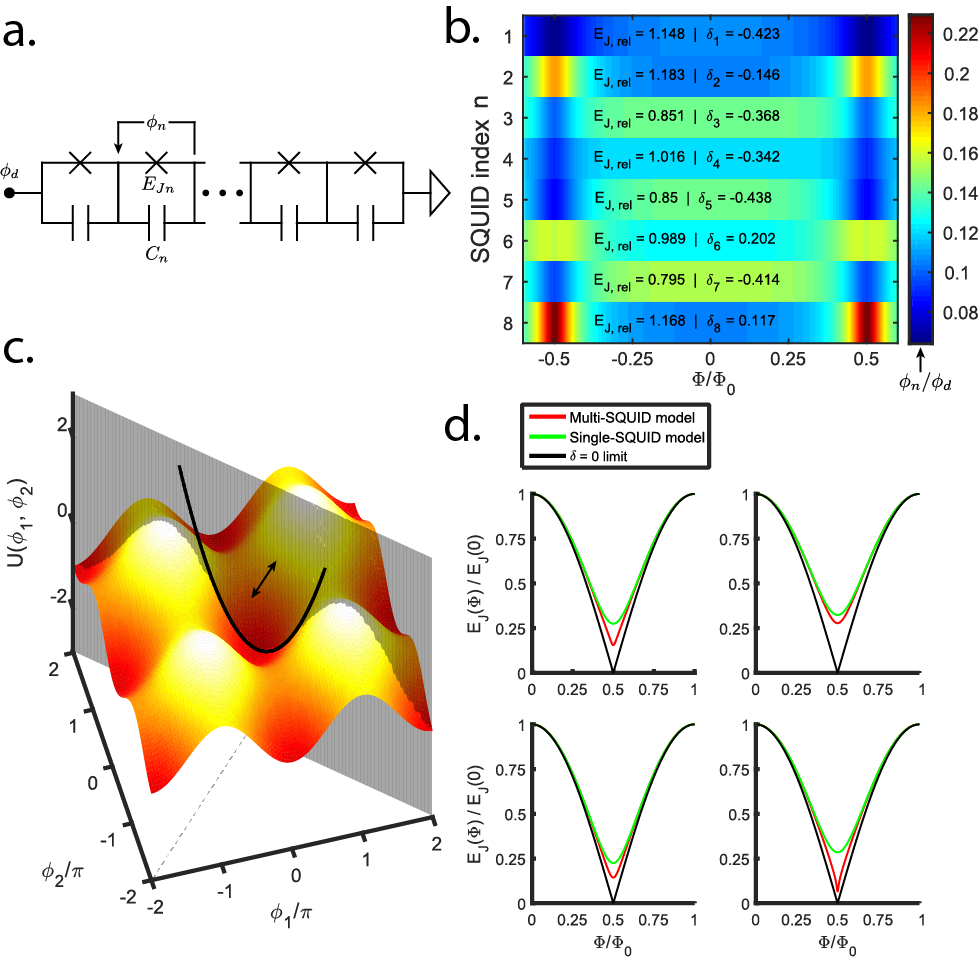}
\caption{\label{fig:squid_array} \textbf{Model for SQUID array}. \bpA Circuit diagram of a SQUID array as a ladder of Josephson junctions and capacitors, in which the twin junctions of each SQUID have been lumped into a single flux-dependent $\EJn$. \bpB Distribution of phases for a randomly-generated array of 8 SQUIDs as a function of flux bias, calculated in the linear voltage-divider approximation in eqns. \ref{eq:voltage_divider_first}-\ref{eq:voltage_divider_last}. Large disorder is used to emphasize that larger phase drops occur across larger inductances, which are designated by smaller $\EJn$ near zero flux, and smaller $\dn$ near half-integer flux.  \bpC Geometric interpretation of the dynamics in eqn. \ref{eq:array_eom_full} for a two-SQUID array. The phase configuration lives on the intersection between the plane and cosine surface, where the plane oscillates along the direction of the dashed line. We approximate the intersection as a parabola confined to the center well and assume the configuration adiabatically follows the bottom of the parabola. \bpD Comparison of $\EJ$ calculations for 4 random 8-SQUID arrays. Red curves represent the full sum in eqn. \ref{eq:EJ_sum_full}, green curves represent the single-SQUID approximation in eqn. \ref{eq:EJ_single_sq}, and black curves represent the zero-disorder limit where both models coincide.}
\end{figure}

The modulated multi-mode Hamiltonian in eqn. \ref{eq:basic_modulated_H} was derived in \cite{Wustmann2013} for a single symmetric SQUID. Here we derive assumptions justifying the treatment of an $\Nsq > 1$ SQUID array as a single SQUID and suggest conditions for which the definition of a single Josephson energy in eqn. \ref{eq:EJdef} is no longer valid. The derivation uses classical phase variables. A similar lumped-element model for arrays of Josephson junction devices is derived in \cite{Frattini2018}.

Given an arbitrary SQUID array, we ask how the phases $\phin$ across each SQUID are distributed while satisfying a parametrically-varying constraint $g(t,\{\phin\} ) = \sum_{n} \phin - \phidcl(t) = 0$. Consider the classical Lagrangian for an array with parametric dependence on flux $f$, adapted from \cite{Wallquist2006, Koch2007}:
\begin{equation} \label{eq:sq_lgrngn_full}
    \lgrngn_{array} = \sum_{n = 1}^{\Nsq} \frac{\rphio^2 \Cn}{2} \phidotn^2 + \EJn(f) \cos \lp \phin - \etan(f) \rp,
\end{equation}

where
\begin{equation} \label{eq:sq_parasitic_phase}
\etan(f) \equiv \arctan(\dn \tan(f)) + \etaon(f).
\end{equation}
The additional phase shift $\etaon(f)$ can arise when the distribution of external flux is asymmetric with respect to the axis of the CPW containing the SQUID array, which is the case for our device where the flux line is parallel to one side of the CPW (see Fig \ref{fig:circuit_diagram}f). For simplicity we treat the full $\etan(f)$ as an unknown small correction. The equations of motion can be derived from the ``augmented Lagrangian'' $\lgrngn_{g} \equiv \lgrngn + \lam(t)g$ where $\lam(t)$ is a Lagrange multiplier; see for example \cite{Boas2006}.
\begin{equation} \label{eq:array_eom_full}
    -\frac{\rphio^2 \Cn}{\EJn} \phiddot_{n} - \sin(\phin - \etan) + \frac{\lam(t)}{\EJn} = 0
\end{equation}
Eqn. \ref{eq:array_eom_full} is equivalent to motion of a particle along the $\Nsq-1$ dimensional intersection of the sinusoidal potential $V = -\sum_{n} \EJn \cos \lp \phin - \etan \rp $ with the vertical plane satisfying $g(t,\{\phin\} ) = 0$. We neglect the acceleration term by noting that in the Fourier domain it equals $\frac{\omega^2}{\omega_{s}^2} \phin[\omega]$, while for small arguments $\sin(\phin - \etan) \approx \phin[\omega] - \etan[\omega]$. $\omega_{s}/\tpi$ is the Josephson plasma frequency of a SQUID, estimated as $21$ GHz in section \ref{sec:flux_tunability}. Assuming the phases follow $\phid(t)$ adiabatically, $\phin[\omega]$ has support only near the populated frequency sites of the resonator, typically $\wn/\tpi \sim 4-6~ \text{GHz}$ such that $\omega^2 / \omega_{s}^2 < 0.1$, (this no longer holds above $7$ GHz). We then write $\lam \approx \EJn \sin(\phin - \etan)$ for all $n$. This assumption is equivalent to pinning the phase configuration to a minimum in the ``intersection potential'' as $\phidcl$ oscillates, however there are an infinite number of such minima and phase slips can occur \cite{Pop2010} when the configuration moves between different wells in the potential $V$. A schematic visualization of the potential landscape for a $2$-SQUID array is shown in Fig. \ref{fig:squid_array}c. For simplicity we assume no phase slips occur ($|\phidcl | \ll 1$ is usually sufficient), and linearize the equations of motion in the potential well closest to $\phin = 0$ for all $n$: $\lam \approx \EJn \lp \cos(\etan) \,  \phin -  \sin(\etan) \rp$. Solving for the phases:
\begin{gather} \label{eq:voltage_divider_first}
    \phin = \xn + \zn \phidcl
    \\
    \zn \equiv \frac{L_{I, n}}{\sum_{m} L_{I, m}}
    \\
    \xn \equiv \tan(\etan) - \zn \sum_{m} \tan(\etam)
    \\ \label{eq:voltage_divider_last}
    L_{I, n} \equiv \frac{\rphio}{\EJn \cos(\etan)}.
\end{gather}
The coefficients $\zn$ represent a voltage divider made of linear inductors, and the $\xn$ represent equilibrium offsets to the quadratic inductor energies due to the phase shifts $\etan$. All coefficients are functions of flux bias $f \equiv \pi \Phi/\Phio$, and an example of the flux dependence $\zn(f)$ for an arbitrarily disordered $8$-SQUID array is shown in Fig. \ref{fig:squid_array}b. The Josephson potential of the $n^{th}$ SQUID is:
\begin{multline} \label{eq:Vn_no_shift}
    \frac{-V_{n}}{\EJn} = \cos [ \tan(\etan) - \etan + \zn ( \phidcl - \sum_{m} \tan(\etam) )].
\end{multline}
We can clean up eqn. \ref{eq:Vn_no_shift} by shifting coordinates to incorporate the DC flux bias $F$ as suggested in \cite{Koch2007}. Returning to eqn. \ref{eq:sq_lgrngn_full}, define $\etan(f) = \etan(F) + \Delta \etan(f)$ where $\Delta \etan$ is a small modulation and we redefine $\phin - \etan(F) \rightarrow \phin$ such that $\phidcl - \sum_{m}\etam(F) \rightarrow \phidcl$. The shift in $\phidcl$ is constant and does not affect the Lagrangian of the coupled CPW as it depends only on spatial and time derivatives of $\phi$ \cite{Eichler13}. Then $\etan$ can be replaced with $\Delta \etan$ in all the expressions of this section and we have
\begin{gather}
    \frac{-V_{n}}{\EJn} = \cos [ \zn(\phidcl - \sum_{m} \Delta \etam + \mathcal{O} (\Delta \etam ^3)) ]
    \\ \label{eq:Vn_simple}
    \frac{-V_{n}}{\EJn(F)} = \cos(\zn(F)\phidcl)
\end{gather}

Eqn. \ref{eq:Vn_simple} neglects the modulated phase shifts $\Delta \etam$ and applies when $\frac{| d_{m} \delta f |}{\cos^2(F)} \ll 1$; in this work we estimate this ratio to be at most $0.015$. The leading-order effect of including $\Delta \etam$ is a classical drive $ \propto \delta f(t) \phidcl$, with the same spectrum as the original modulation signal and typically far off-resonant from all sites.

An equivalent single-SQUID $\EJ$ can be defined using the inductive current through the SQUID array:
\begin{multline}
        \rphio I_{ind} = \partial_{\phidcl} \lgrngn_{\text{array}} 
        \\
        = \sum_{n} \EJn \zn \sin(\zn \phidcl) \approx \sum_{n} \EJn \zn^2 \phidcl \equiv \EJ \phidcl
\end{multline}
Using the voltage-divider coefficients $\zn$ we can confirm $L_{J} \equiv \rphio^2/\EJ = \sum_{n} \LJn$. Writing the explicit flux dependence of the coefficients and using a series of Taylor expansions:
\begin{multline} \label{eq:EJ_sum_full}
    \EJ^{-1}(f) = \sum_{n} \lp E_{J0, n} |\cos(f)| \sqrt{1 + \dn^2 \tan^2(f)} \rp ^{-1}
    \\
    \approx \Nsq|\cos(f)|^{-1} \frac{\langle \LJo \rangle }{\rphio} \lp 1 - \half \frac{\langle \LJo d^2 \rangle }{\langle \LJo \rangle } \tan^2(f) \rp
\end{multline}
\begin{gather} \label{eq:EJ_single_sq}
    \EJ(f) \approx \frac{\rphio^2}{\Nsq \langle \LJo \rangle } \sqrt{\cos^2(f) + \dsq^2 \sin^2(f)}
    \\
    \dsq^2 \equiv \frac{\langle \LJo d^2 \rangle }{\langle \LJo \rangle} = \langle \zn(f=0)d^2 \rangle
\end{gather}
The effective asymmetry parameter $\dsq^2$ is obtained by averaging over the voltage-divider distribution at zero applied flux, i.e. the minimum Josephson inductances. Eqn. \ref{eq:EJ_single_sq} is valid only where $\dn^2 \tan^2(f) \ll 1$, which for our experimental conditions $|\dn| < 0.1$ and $ |f/\pi \mod 1| < 0.33$ is satisfied as $\dn^2 \tan^2(f) < 0.03$. The expansion cannot be used near the tangent poles at $|f/\pi \mod 1| = 0.5$, as seen in Fig. \ref{fig:squid_array}d where the effective $\EJ$ of disordered arrays (red) dips below the single-SQUID approximation (green). A similar argument can be used to define a single-SQUID, flux-independent capacitance using the approximations $\Delta \etam \approx 0$ and $\partial_{t} \zn \approx 0$. To summarize, our treatment of a SQUID array with the single-SQUID model requires $\omega^2 / \omega_{s}^2 \ll 1$ (resonator excitation well below the Josephson plasma frequency), no phase slips, and $\dn^2 \tan^2(f) \ll 1$ (flux bias kept within the range where disorder is perturbative).

\section{Equivalence of transient spectra and scattering parameters} \label{app:transient_scattering_equivalence}
Scattering parameters between ports in frequency space can be estimated from the Fourier transform of transient data, equivalent to eqn. \ref{eq:scattering_with_input_off} in the main text. Consider an initial vector of coherent states $\betaul(t = 0^+)$ that undergoes linear free evolution $\dot{\betaul} = -i \Hbold \betaul$ (H is here a classical matrix which may have a non-Hermitian part to describe loss). This linear evolution is independent of the exact way $\betaul(t = 0^+)$ was prepared, so we can choose any convenient model for the input signal so long as we only consider measurements after the input signal is turned off (i.e. $t > 0$). Suppose our model excites site $M$ with an impulse at $t = 0$. The impulse must have finite duration to confine its spectrum about site $M$; we revisit this point after a simpler discussion using $\delta(t)$:
\begin{equation} \label{eq:impulse_evol}
    \betadotm = -i \sum_{n} \Hbold_{mn} \betan + \sqrt{\kapm^{e}} B_{in} \delta(t) \delta_{mM}
\end{equation}

Integrating over a symmetric interval about $t = 0$, taking its width to zero and assuming $\betam(t<0) = 0$:
\begin{equation} \label{eq:impulse_init}
    \betam(t = 0^+) = \lim_{\eps \rightarrow 0^+} \int_{-\eps}^{\eps} \betadotm dt = \sqrt{\kapm^{e}} B_{in} \delta_{mM}
\end{equation}

We can use eqn. \ref{eq:impulse_init} to model any initial amplitude. To model the collection of data for $t>0$, consider a step filter applied to the output signal:
\begin{equation}
    \betaout(t) \thplus(t) \equiv \left\{\begin{matrix}
    0 & t <= 0
    \\
    \betaout(t) & t > 0
\end{matrix}\right.
\end{equation}
\begin{multline}
    (\betaoutm \thplus) \ofw = \fsymco \int_{0^+}^{\infty} \betaoutm(t) e^{i \omega t} dt
    \\
    = - \sqrt{\frac{\kapm^{e}}{\tpi}} \int_{0^+}^{\infty} \betam(t) e^{i \omega t} dt 
    \\ 
    = - \sqrt{\kapm^{e}} (\betam \thplus) \ofw
\end{multline}
Multiplying eqn. \ref{eq:impulse_evol} by the step filter and integrating by parts (essentially a Laplace transform with $s = -i\omega$):
\begin{multline}
    \int_{0^+}^{\infty} \betadotm(t) e^{i \omega t} dt = -i\omega \int_{0^+}^{\infty} \betam(t) e^{- \omega t} dt - \betam(0^+) 
    \\
    = -i \sum_{n} \Hbold_{mn} \int_{0^+}^{\infty} \betan(t) e^{i \omega t} dt 
\end{multline}
Rewriting with filtered transforms and site vectors:
\begin{equation}
    -i \sum_{n} (\omega \delta_{mn} - \Hbold_{mn}) (\betam \thplus) \ofw = \betam(0^+)
\end{equation}
\begin{equation}
    (\beta \thplus) \ofw = i(\omega \boldsymbol{I} - \Hbold)^{-1} \sqrt{\kappa_{M}^{in}} B_{in} \underline{u}_{M}
\end{equation}
\begin{multline} \label{eq:scattering_transient_equiv}
    (\betaout \thplus) \ofw = -i \sqrt{\kapemat} (\omega\boldsymbol{I} - \Hbold)^{-1} \sqrt{\kapemat} B_{in} \underline{u}_{M}
    \\
    = (\Sbold \ofw - \boldsymbol{I}) B_{in} \underline{u}_{M}
\end{multline}
The second line of eqn. \ref{eq:scattering_transient_equiv} follows from comparison to the steady-state scattering matrix in eqn. \ref{eq:scattering_with_input_on}, where loss is incorporated into $\Hbold$. From this we find $(\Sbold_{mn} \ofw - \delta_{mn}) \propto (\betaoutm \thplus) \ofw $ when site $n$ is initially excited, as in Figs \ref{fig:spectra} and \ref{fig:single_site_ex}. While more general form of eqn. \ref{eq:scattering_transient_equiv} can be derived for a multi-site initial excitation, the given form is convenient for measuring scattering parameters.

Returning to a finite-impulse model, we need the pulse bandwidth to be much larger than the site bandwidth ($W_\text{pulse} \gg W_\text{site}$) so the input spectrum is locally constant, but smaller than the distance between neighboring sites ($W_\text{pulse} < 2 \Wmod - W_\text{site}$), where we take the sites to be well-resolved ($W_\text{site} \ll \Wmod$). The corresponding pulse $\delta_{\tau}(t)$ is appreciable for $t \in (-\tau/2, \tau/2)$, with carrier frequency $\wm$ centered on the target site. To estimate the size of $\tau$, take $W_{pulse} = \Wmod$ (pulse spectrum extends halfway to neighboring sites), and use $W_{pulse} \tau  > \tpi$ such that $\tau  > 7 ns$ and we take $\tau^+/2$ as the ``initial'' time instead of $0^+$. Because $\tau \ll J_{typ}^{-1}$, the system dynamics are nearly stationary during the pulse and the finite width has minimal effect on the model.

\section{Calibrating transient scattering gain} \label{app:transient_scattering_gain}
The spectra in Fig. \ref{fig:spectra}c are normalized using a procedure adapted from \cite{Ma2017a}. For a lattice system with linear couplings, the normalized single-port reflection coefficients satisfy
\begin{equation} \label{eq:kappa_e_integral}
   \Pr \intallreal \Sbold_{mm}\ofw d\w = \pi \kapm^{e}
\end{equation}

where $\Pr$ is the Cauchy principal value and $\Sbold_{mn}\ofw \equiv \Sbold_{mn}^\text{full}\ofw - \delta_{mn}$ are scattering parameters with the component from direct reflection subtracted, as in eqn. \ref{eq:scattering_transient_equiv}. As argued above, we expect the Fourier transform of the transient output voltage to be proportional to $\Sbold_{mn} \ofw$ when site $n$ is initially excited and output from site $m$ is read out. However we need to account for the gain and loss in the components (cables, filters, amplifiers, and mixers) on the input and output sides of the device. We use a simple model with a constant initial amplitude $\vo$ and one frequency-dependent gain on each side:
\begin{multline} \label{eq:gain_model_full}
    v_{mn} \lb \wmread + \w , \wnex \rb =
    \\
    \Gmout\lb \wmread + \w \rb \Sbold_{mn} \ofw \Gnin \lb \wnex \rb \vo
\end{multline}
We need to be careful about how these frequencies are defined. The readout frequencies specify the rotating frame used throughout this work, $\wmread \equiv \wm' \equiv \wo + m\Wmod$, where the same $\wo$ and numbering convention for $m$ are used for all input and output sites. The excitation frequencies $\wnex \equiv \wn$ are chosen as the center frequencies of the \textit{uncoupled} modes $n$, which differ from the readout frequencies by the disorder $\Deln$. Eqn. \ref{eq:gain_model_full} is labeled with redundancy to illustrate that the input and output gains are functions of lab-frame frequency, $\w$ is the frequency coordinate of the output spectrum about site $m$, and the input signal is assumed to be monochromatic while the output signal has a bandwidth of a few MHz about each site. Explicitly moving to the rotating frame and writing both off-diagonal and diagonal site relations:
\begin{gather} \label{eq:gain_mn_rot}
    v_{mn}\lb \w, \Deln \rb  = \vo \Gmout \ofw \Sbold_{mn} \ofw \Gnin \lb \Deln \rb 
    \\ \label{eq:gain_nm_rot}
    v_{nm}\lb \w, \Delm \rb  = \vo \Gnout \ofw \Sbold_{nm} \ofw \Gmin \lb \Delm \rb 
    \\ \label{eq:gain_nn_rot}
    v_{nn}\lb \w, \Deln \rb  = \vo \Gnout \ofw \Sbold_{nn} \ofw \Gnin \lb \Deln \rb 
\end{gather}
Combining eqns. \ref{eq:gain_mn_rot}-\ref{eq:gain_nn_rot} we find an expected result:
\begin{equation}
    \frac{\Sbold_{mn} \ofw \Sbold_{nm} \ofw}{ \Sbold_{mm} \ofw \Sbold_{nn}\ofw} = \frac{v_{mn} [\w, \Deln] v_{nm} [\w, \Delm] }{ v_{mm} [\w, \Delm] v_{nn} [\w, \Deln]}
\end{equation}

Next we define $G_{mm} \ofw \equiv \vo \Gmout \ofw \Gmin \lb \Delm \rb$ so that eqn. \ref{eq:kappa_e_integral} can be written as (suppressing $\Delm$)
\begin{equation}
    \pi \kapm^{e} = \Pr \intallreal \frac{ v_{mm} \ofw }{G_{mm} \ofw} d\w.
\end{equation}
Our next simplifying assumption is that $G_{mm} \ofw \approx G_{mm}$ can be treated as constant over the bandwidth of site $m$ ($< 10 $ MHz), in which case we can calculate
\begin{equation}
    G_{mm} \approx \frac{\Pr \intallreal  v_{mm} \ofw d\w}{\pi \kapm^{e}}.
\end{equation}
The scattering amplitudes plotted in Fig. \ref{fig:spectra} are calculated by assuming reciprocity in the synthetic dimension, i.e. $\Sbold_{mn} \ofw =  \Sbold_{nm} \ofw$ where there are no magnetic fields or other gauge fields present \cite{Pozar2012}. This assumption is theoretically valid when only one modulation frequency is used.
\begin{equation}
    ``\Sbold_{mn} \ofw'' \equiv \lbar \sqrt{\Sbold_{mn} \ofw \Sbold_{nm} \ofw} \rbar = \lbar \sqrt{\frac{v_{mn} \ofw v_{nm} \ofw }{ G_{mm} G_{nn}}} \rbar.
\end{equation}
The above assumptions imply that $v_{mn} \ofw/v_{nm}\ofw$ should be constant across a site bandwidth, however this is not exactly validated by the experimental spectra. The estimated ``$\Sbold_{mn} \ofw$'' is therefore more useful qualitatively, and a more robust calibration of the input and output lines is needed to perform calculations of matrix elements $\Hbold_{mn}$ as proposed in \cite{Ma2017a}.

\section{Nearest-neighbor interference} \label{app:neighbor_interf_deriv}

\begin{figure}
\includegraphics{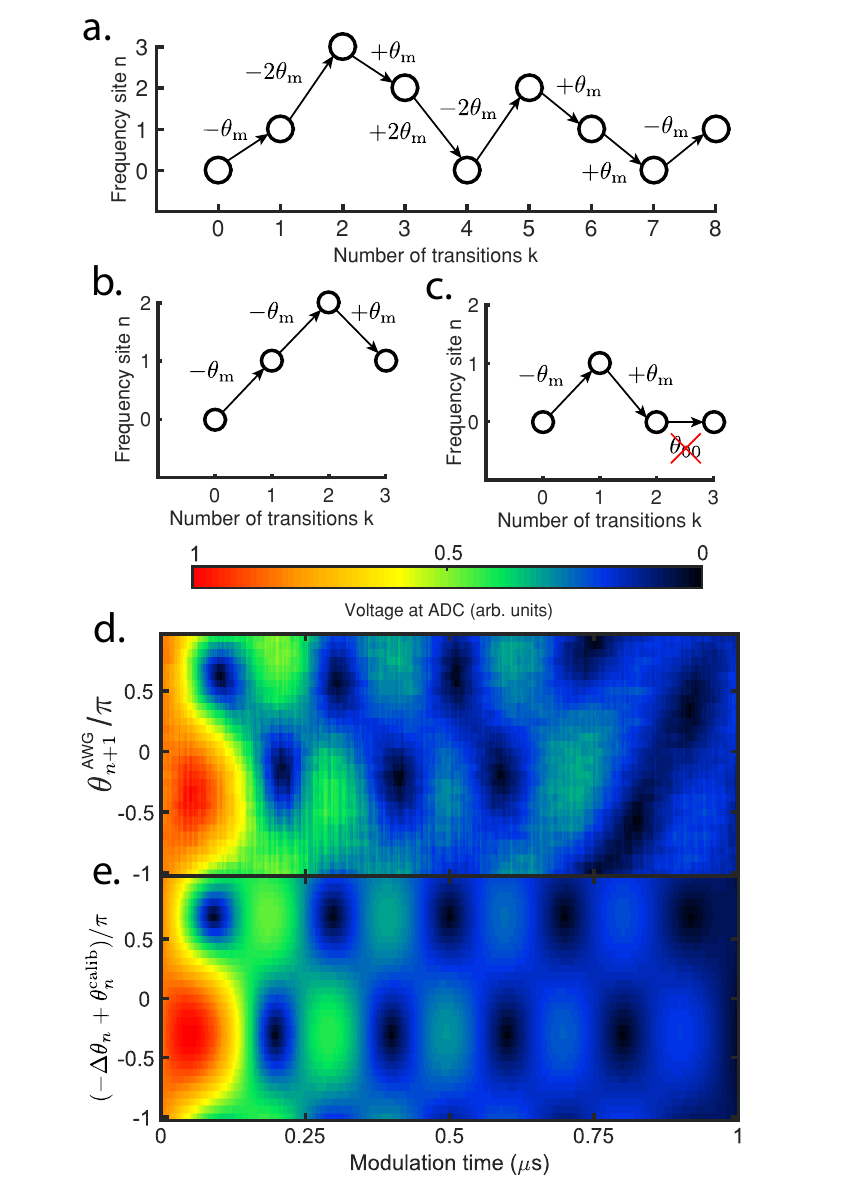}
\caption{\label{fig:neighbor_interf} \textbf{Diagrams and extended-time behavior for nearest-neighbor interference.} \bpA Example off-diagonal diagram satisfying eqn. \ref{eq:off_diag_diagr}, in which the product of all propagation phases depends only on the distance between the initial and final sites. \bpB Example diagram connecting $n = (0, 1)$ with only single hops such that the total number of transitions is odd; this type of diagram is included in deriving eqn. \ref{eq:neighbor_interf_2}. \bpC Example diagram connecting $n = (0, 0)$ with an odd number of hops by using a diagonal matrix element; we ignore such diagrams at short times. (\textbf{D, E}) Experimental and calculated voltage magnitudes over 1 $\mus$ for initial sites $n = (-1, 0)$, with output measured at $n = -1$. The first $0.25~ \mus$ of \bpD are reproduced in Fig. \ref{fig:multi_site_ex} of the main text. The calculation in \bpE assumes a uniform coupling rate $J/\tpi = 1.25~ \text{MHz}$, uniform loss $\kappa/\tpi = 90~ \text{kHz}$, and a calibration phase $\theta^\text{calib}_{n = -1}  = -0.20 \pi$. This calibration phase absorbs the modulation phase for simpler plot labeling.}
\end{figure}

This section derives eqn. \ref{eq:neighbor_interf_1} and provides an argument for a sinusoidal interference pattern between two neighboring light cones at short modulation times. Starting from the time evolution at site $n$ and expanding the vector of initial amplitudes,
\begin{equation} \label{eq:propagator_mat_el}
    \betan(t) = \udagn e^{-i \Hbold t} \betaul(0) = \sum_{m} \betam(0)  \udagn e^{-i \Hbold t} \um.
\end{equation}
we expand the matrix element $(e^{-i \Hbold t})_{nm}$ as a Dyson series; time-ordering is trivial because $\Hbold$ has no explicit time-dependence in the rotating-wave approximation.
\begin{multline} \label{eq:propagator_dyson}
    \udagn e^{-i \Hbold t} \um = \sum_{k} \frac{(-it)^{k}}{k!} \udagn \Hbold^{k} \um 
    \\
    = \sum_{k} \frac{(-it)^{k}}{k!} \sum_{p, q, ..., v} \Hbold_{np} \Hbold_{pq} \cdot \cdot \cdot \Hbold_{vm}
\end{multline}
The second line of eqn. \ref{eq:propagator_dyson} motivates the use of diagrams to understand propagation from site $m$ to $n$ as the sum of all possible ``hopping'' trajectories on the lattice, as in Fig. \ref{fig:neighbor_interf}a. For an ideal lattice-translation experiment there exists a set of off-diagonal matrix elements which are much larger than the diagonal elements, so we focus on ``off-diagonal'' diagrams that contain only terms like $\Hbold_{n, p \neq n}$. When only one modulation frequency is used, all off-diagonal elements satisfy $\Hbold_{np} = |\Hbold_{np}| e^{-i(n-p)\thmod}$ and therefore
\begin{multline} \label{eq:off_diag_diagr}
    \udagn e^{-i \Hbold t} \um \mid_\text{off-diag} =
    \\
    \sum_{k} \frac{(-it)^{k}}{k!} \sum_{p\neq q \neq ... \neq v} |\Hbold_{np}| |\Hbold_{pq}| \cdot \cdot \cdot |\Hbold_{vm}|e^{-i(n-m)\thmod}.
\end{multline}
We note that each term in eqn. \ref{eq:off_diag_diagr} equals a real coefficient times $i^k e^{-i(n-m)\thmod}$.  Focusing on the calibration experiment in Fig. \ref{fig:multi_site_ex}, we consider an initial state $\beta(0) = \un + r e^{i \Delta\theta_{n}} \underline{u}_{n+1}$ and which allows us to probe the matrix elements $(e^{-i \Hbold t})_{nn}$ and $(e^{-i \Hbold t})_{n,n+1}$. For nearest-neighbor coupling all off-diagonal diagrams connecting site $n$ to itself contain an even number of ``hops'' i.e. even $k$, while all diagrams connecting $n$ to $n+1$ contain an odd number of hops (see illustration in Fig. \ref{fig:neighbor_interf}b,c). Factoring out possibly-complex quantities we obtain
\begin{eqnarray}
    (e^{-i \Hbold t})_{nn} \mid_\text{off-diag} &=& \Un (t) \nonumber\\
    (e^{-i \Hbold t})_{n,n+1} \mid_\text{off-diag} &=& -i e^{i\thmod}   \Unp (t)
\end{eqnarray}
We can then write
\begin{equation}
    \betan (t)\mid_\text{off-diag}  = \Un (t) - ire^{i(\Delta\theta_{n} + \thmod)}\Unp (t)
\end{equation}
resulting in
\begin{multline} \label{eq:neighbor_interf_1_copy}
    |\beta_{n}(t)|^2\mid_\text{off-diag}  = |\Un(t)|^2 + r^2 |\Unp(t)|^2 
    \\
    - 2 \text{Re} \lb i r  \Un(t)^{*} \Unp(t) e^{i(\Delta\theta_{n} + \thmod)}\rb
\end{multline}
which reproduces eqn. \ref{eq:neighbor_interf_1} in the main text. At short modulation times when $\mathcal{U}_{n}$ and $\mathcal{U}_{n+1}$ are approximately real-valued, eqn. \ref{eq:neighbor_interf_1_copy} simplifies to:
\begin{multline} \label{eq:neighbor_interf_2}
    |\betan (t)|^2\mid_\text{off-diag}  = \Un (t)^2 + r^2 \Unp (t)^2 + 
    \\
    2r \Un (t) \Unp (t) \sin(\Delta\theta_{n} + \thmod).
\end{multline}
At later times an additional time-dependent phase appears in the sine. To first order in $\Hbold t$ we have $\Un(|\Hbold t \ll 1|)  \approx 1$ and $\Unp(|\Hbold t \ll 1|) \approx |J_{n, n+1}|t$, so the coefficient of the interference term is positive for all $n$ at small $t$ regardless of whether the coupling rates are translationally-invariant. We therefore have an interference pattern for each pair of sites ($n, n+1$) that can be used to map the relative driving phase at the AWG to the on-chip phase $\Delta\theta_{n} + \thmod$.

In deriving eqn. \ref{eq:neighbor_interf_2} we ignored diagrams with diagonal matrix elements 
\begin{equation}
    \Hbold_{nn} = \Deln - i \kapn = \sqrt{\Deln^2 + \kapn^2}e^{-i \tan^{-1} \lp \frac{\kapn}{\Deln} \rp}
\end{equation}
which introduce phases unrelated to $\thmod$ and allow both even and odd diagrams between sites ($n, n+1$) and $n$, with a net effect of adding a time-dependent phase to the sine in eqn. \ref{eq:neighbor_interf_2}. This is easiest to justify for short times such that $\sqrt{\Deln^2 + \kapn^2} t \ll 1$, which is compatible with the off-diagonal calculation's short-time limit $|J_{n, n+1}|t < 1$ when $\sqrt{\Deln^2 + \kapn^2} \ll  |J_{n, n+1}|$. In this work a typical ratio is $\frac{\sqrt{\Deln^2 + \kapn^2}}{|J_{n, n+1}|} \sim 0.1$.

Fig. \ref{fig:neighbor_interf}(d, e) compares experimental interference data with a theoretical calculation using uniform coupling and loss rates. Voltage magnitudes are plotted without squaring to show better contrast in the oscillations at later times. The calculation shows a sinusoidal relationship with a fixed phase along to the $\theta$-axis for at least $1~\mus$, while experimental data shows a sinusoid that drifts slowly upward for the first $0.6 \mus$ and then rapidly deviates from the calculation. This deviation suggests that higher-order diagrams involving diagonal matrix elements can no longer be neglected as $t$ approaches $0.6\mus$, and we limit our phase-calibration analysis to $t < 0.15 \mus$.

\section{Second-nearest-neighbor coupling} \label{app:second_nearest_coupling}

\begin{figure}
\includegraphics{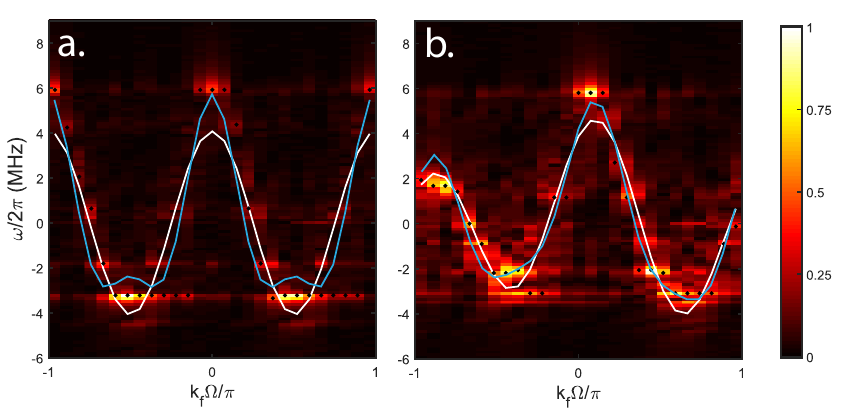}
\caption{\label{fig:second_nearest_neighbor} \textbf{Fourier transforms for second-nearest-neighbor coupling}. \bpA Single-tone modulation at 310.2 MHz showing two symmetric dispersion periods. White and blue curves are fits to models incorporating the lowest and two lowest modulation harmonics, respectively.  \bpB Two-tone modulation at 155.1 and 310.2 MHz showing asymmetric dispersion. The white curve is a fit incorporating only the lowest harmonic of each tone; the blue curve includes the second harmonic of the $2\Wmod$ tone as in \bpA. Black dots overlaid on both plots show the maximum amplitude for each value of $\kf \Wmod$.}
\end{figure}

Flux modulation at frequencies near twice the average FSR drives coupling between second-nearest neighbors with rates $J_{m, m+2}$. Using methods described in sections (\ref{sec:transient_prop} and \ref{sec:multi_site}) we observe coupling between second-nearest-neighbors and extract approximate dispersion relations. For these experiments, 4 dB of attenuation was removed from the AWG Channel 2 output, providing additional amplitude for two-tone modulation. The data in Fig. \ref{fig:second_nearest_neighbor} display qualitative features of higher-order coupling, though quantitative fits are poorer than for nearest-neighbor coupling. Fig. \ref{fig:second_nearest_neighbor}a contains experimental data for single-tone modulation at $2\Wmod/\tpi = 310.2~ \text{MHz}$, with least-squares fits to the following model:
\begin{equation}
    \omega(k) = 2 |J_{2}| \cos(2 \kf \Omega) + 2 |J_{4}| \cos(4 \kf \Omega)
\end{equation}
where $\Wmod/\tpi = 155.1~ \text{MHz}$ and the $J_{4}$ term arises from the second modulation harmonic, assumed in-phase with the first harmonic. The white curve assumes $J_{4} = 0$ and fits $|J_{2}|/\tpi = 2.04~ \text{MHz}$, while the blue curve fits $|J_{2}|/\tpi = 2.04~ \text{MHz}$ and $|J_{4}|/\tpi = 0.835~ \text{MHz}$. The two-parameter for captures the main qualitative features (symmetry about $\kf \Wmod = 0$, sharper peaks, wider troughs) better than the one-parameter fit, suggesting the second harmonic of the modulation tone is non-negligible. This is consistent with reduced attenuation in the modulation line, since a larger flux modulation $\sim 0.1 \Phio$ begins to access the curvature of the SQUID-array $\EJ$ in eqn. \ref{eq:EJdef}.

Fig. \ref{fig:second_nearest_neighbor}b contains data for two-tone modulation at $(1,2)\Wmod/\tpi = (155.1, 310.2)~ \text{MHz}$, with a least-squares fit to:
\begin{multline}
    \omega(k) = 2 |J_{1}| \cos(\kf \Omega) + 2 |J_{2}| \cos(2 \kf \Omega + \theta_{2})
    \\
    + 2 |J_{4}| \cos(4 \kf \Omega + 2\theta_{2}) 
\end{multline}

The two-tone drive introduces a gauge-invariant phase $\theta_{2} \approx -0.08 \pi$ and asymmetric dispersion about $\kf \Wmod = 0$, which can be interpreted in terms of a synthetic gauge field as in Ref. \cite{Dutt2019}. White and blue curves are fits assuming zero and nonzero $J_{4}$, respectively. In both cases we find $|J_{1}|/\tpi = 0.659~ \text{MHz}$ and $|J_{2}|/\tpi = 1.71~ \text{MHz}$; for the blue curve $|J_{4}|/\tpi = 0.399~ \text{MHz}$ which leads to a better qualitative fit. The second harmonic of the 155.1 MHz tone contributes another $J_{2}$-like term with a different phase, but is neglected in the fits as we expect its coupling rate to be below 100 kHz.

\section{Time-reversed coupling and parasitic oscillations} \label{app:time_rev_coupling}

\begin{figure}
\includegraphics{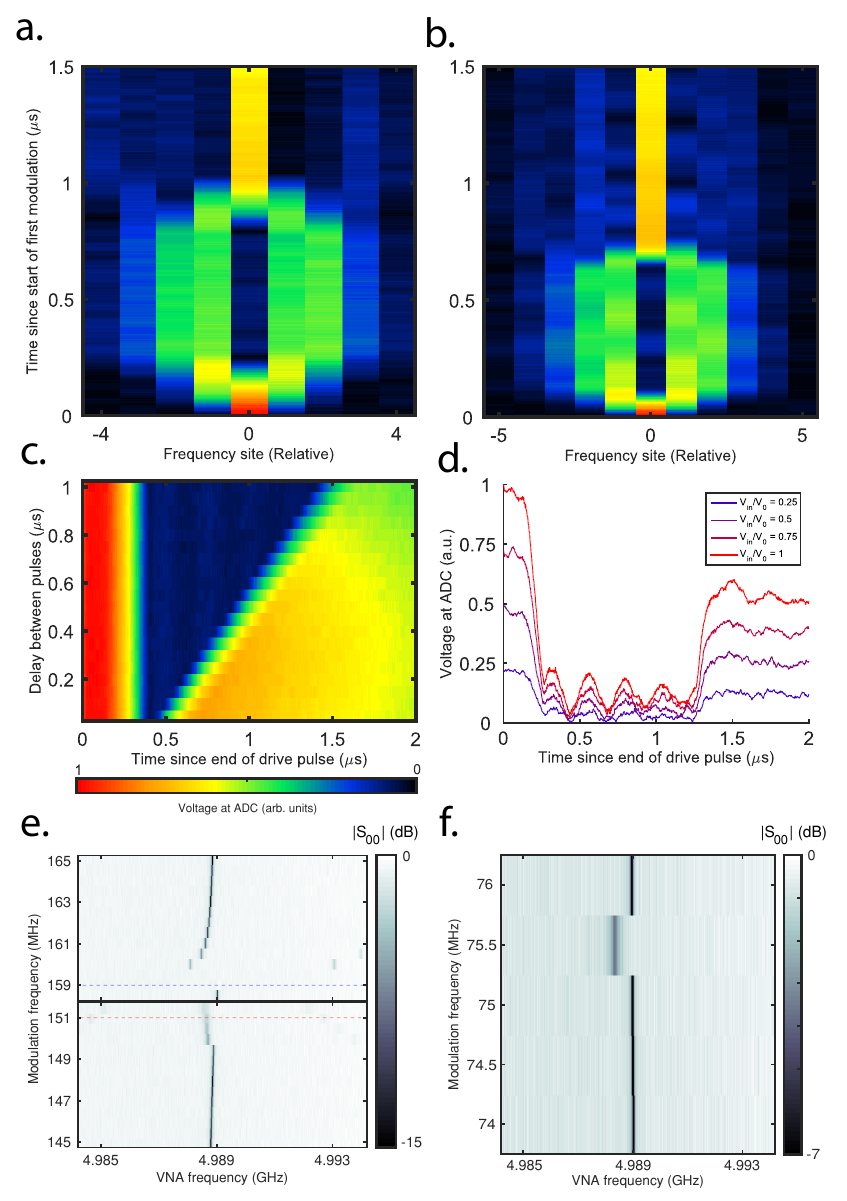}
\caption{\label{fig:coupling_reversal} \textbf{Experimental data for time-reversed coupling and parasitic oscillations}. (\textbf{A, B}) Multi-site output traces for modulation pulses at amplitudes of $0.062 \Phio$ and $0.031 \Phio$ respectively; all traces are smoothed by a 16-point moving average. Pulse durations are chosen to minimize the amplitude remaining in site 0 during the delay between pulses (here 0.5 $\mus$). \bpC Single-site output traces at n = 0 as the delay is swept from 0.05 to 1 $\mus$, with flux amplitude $0.031 \Phio$. \bpD Single-site output traces at n = 0 for varying \textit{resonator excitation} amplitude and flux amplitude $0.062 \Phio$, showing 4 periods of parasitic oscillation within the 1 $\mus$ pulse delay. \bpE Single-site reflection spectra at n = 0, sweeping modulation frequency at larger detuning. Dashed lines indicate approximate modulation frequencies that induce Bloch oscillations at 4 MHz. \bpF Single-site reflection spectra at n = 0, sweeping modulation frequency near the expected lowest harmonic of the device. A large redshift is seen for modulation at 75.5 MHz, which we attribute to a cross-Kerr interaction when the lowest harmonic is directly excited by the flux line.}
\end{figure}

Consider the tight-binding Hamiltonian separated into diagonal and off-diagonal elements in the frequency-site basis: $\Hhat = \Hhat_{0} + \Hhat_{\{J\}}$. Time evolution is given by 
\begin{equation}
    \hat{U}(t) = e^{-i\Hhat t} = e^{-i(-\Hhat)(-t)},
\end{equation}
which suggests an equivalence between time-reversal and instantaneously flipping the signs of every term in the Hamiltonian. Recalling that off-diagonal elements satisfy $\Hbold_{np} = |\Hbold_{np}| e^{-i(n-p)\thmod}$ for a single modulation frequency near the FSR, we can introduce an time-reversal for odd $(n-p)$ by applying a phase shift $\thmod \rightarrow \thmod \pm \pi$, however for even $(n-p)$ the matrix element is invariant. A typical situation where this applies is nearest-neighbor coupling ($|n-p| = 1$, time-reversible) in the presence of on-site disorder and loss ($n-p = 0$, not reversible). We observe this partial time-reversal and utilize it to create a tunable output delay in site $n = 0$ as in Fig. \ref{fig:coupling_reversal}. After driving site 0 to steady-state, a short modulation pulse spreads the excitation almost entirely into the nearby sites. After a tunable delay time, a second modulation pulse with the same amplitude and duration causes the excitation to return to the initial site. The key point in this schematic is choosing the phase of the second modulation pulse, which must be advanced by $\pi$ with respect to the phase the first pulse would have \textit{if it had been left on continuously}. We scale the pulse duration inversely with pulse amplitude, as in Figs. \ref{fig:coupling_reversal}a (Half flux amplitude, $272$ ns pulses) and \ref{fig:coupling_reversal}b (Maximum flux amplitude, $136$ ns pulses)

Leakage of amplitudes into sites $n \neq 0$ occurs during the reversal pulse, which we anticipate even when the coupling phases are exactly reversed. This is due to on-site disorder, which distorts the distribution of phases on the lattice after sufficient delay time and cannot be time-reversed by this scheme. Empirically we find this leakage to be more severe at larger modulation amplitudes and short pulse durations. In these cases we also observe small oscillations in the output amplitudes when the modulation is \textit{off}, which is not accounted for by the model presented in this work, even with a disordered Hamiltonian. Oscillations at neighboring sites appear to be out of phase in Fig. \ref{fig:coupling_reversal}b, suggesting that excitations are still exchanging between sites after the first modulation pulse is off. The time scale of the oscillations appears to be independent of the power in the resonator, as in \ref{fig:coupling_reversal}d where oscillations at $\sim 4~\text{MHz}$ occur for different amplitudes of resonator excitation. The envelope of these oscillations resembles Bloch oscillations more strongly than resonant coupling (compare Fig. \ref{fig:single_site_ex}(g,d)), which implies parasitic modulation near $155.1 \pm 4~ \text{MHz}$. The modulation pulses in this experiment have a sinc spectrum where the bandwidth of the primary lobe is nearly 15 MHz, large enough to directly excite resonances in the range $155.1 \pm 4~ \text{MHz}$ if they exist. Parasitic modulation could then occur if the excited fields leak into the SQUID array. 

To investigate this parasitic modulation, we measure single-site reflection spectra at $n = 0$ while sweeping the frequency of a continuous modulation tone. The mode redshifts and broadens as the modulation frequency approaches 151 or 159 MHz, and at 159 MHz disappears entirely (Fig. \ref{fig:coupling_reversal}e). We interpret these shifts as quartic-order nonlinear interactions, possibly cross-Kerr terms of form $-\chi_{mn} \bdagm \bmnew \bdagn \bn$ which cause a redshift of mode $m$ based on the Fock occupation of mode $n$ (or vice-versa). Terms of this type survive the rotating-wave approximation for all pairs $(m,n)$. When mode 0 is driven weakly as in \ref{fig:coupling_reversal}e, a cross-Kerr redshift indicates large excitation of a different mode. We suspect the shift at 151 MHz is related to parametric-oscillator pumping in the lowest harmonic of the device, $n = -32$, which we expect to lie near 75 MHz. This excitation scheme uses the interaction $\hat{b}_{-32}^2 + (\hat{b}_{-32}^{\dag})^2$, which excites pairs of photons. We attempt to drive the lowest harmonic directly, and observe another redshift in mode $n = 0$ when the modulation frequency is 75.5 = 151/2 MHz (Fig. \ref{fig:coupling_reversal}f). The shift is consistent with an interaction $-\chi(0, -32)\hat{b}^{\dag}_{0} \hat{b}_{0} \hat{b}^{\dag}_{-32} \hat{b}_{-32}$ assuming the lowest harmonic is excited through the mutual inductance of the flux line and resonator. However only the second harmonic of the 75.5 MHz modulation tone would cause 4 MHz Bloch oscillations, so leakage of the $n = -32$ field into the SQUID array would need to be very large. The parasitic excitation at 159 MHz may provide a better explanation than 151 MHz, though we have not identified a possible cause and we leave rigorous characterization for a future work.

\section{Elastic scattering from a frequency defect} \label{app:elastic_scattering_freq}

\begin{figure}
\includegraphics{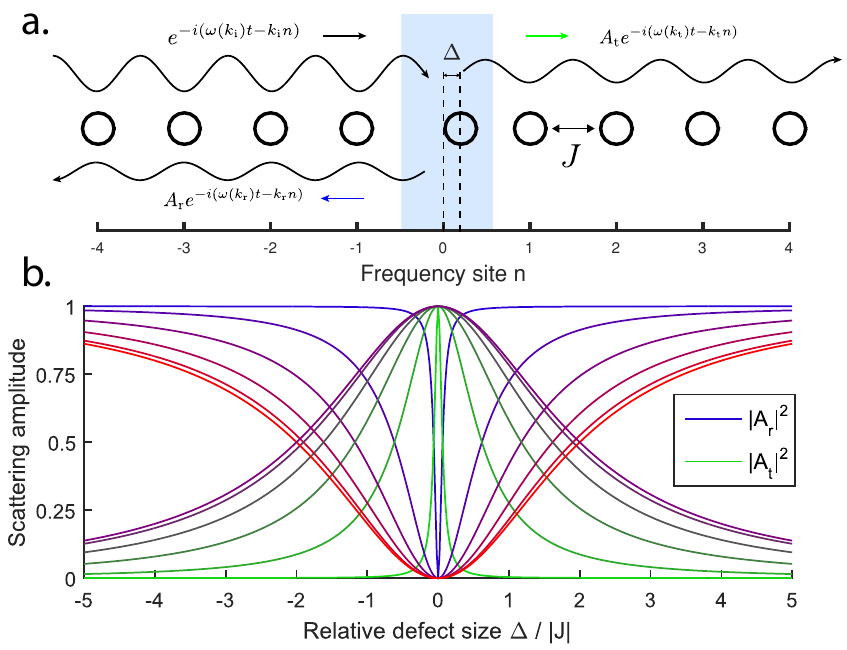}
\caption{\label{fig:freq_defect} \textbf{1D scattering model} \bpA Schematic for reflected and transmitted waves when an incident wave approaches a single defect from the left. Uniform nearest-neighbor coupling and zero on-site energy outside the defect are assumed. \bpB Amplitudes of reflected and transmitted waves as a function of defect size and effective wavevector $k_\text{eff} \equiv \ki + \thJ$. Overlaid plots display $|\Ar|^2$ (blue-to-red gradient) and $|\At|^2$ (green-to-purple gradient) as $|k_\text{eff}|$ varies from $0.01\pi$ to $0.5\pi$.}
\end{figure}

A major constraint in this work is the presence of ``barrier sites'' where large detuning and/or loss causes reflection of propagating lattice states. We derive a simple model for scattering at a site with detuning $\Delta$, on an otherwise translationally-invariant 1D lattice. We neglect loss rates $\kappa$ as a first approximation in the strong-coupling limit where $J \gg \kappa$. In analogy to standard 1D scattering problems \cite{Shankar1994} we postulate solutions on either side of the defect that are separately eigenstates of the defect-less Hamiltonian, and stitch these solutions together with a boundary condition at the defect. We assume an incident and reflected wave to the left of the defect ($n < 0$) and a transmitted wave to the right ($n > 0$) as shown in Fig. \ref{fig:freq_defect}a; for a linear Hamiltonian the problem has arbitrary scaling so the incident wave is assigned an amplitude of 1 without loss of generality. We suppress operator hats and focus on classical waves; quantization can be restored by multiplying the final expression by the incident-wave operator $\hat{A}_\text{i}$. 
\begin{gather} \label{eq:a_n_minus}
    a_{n_{-}} = e^{-i\lp\w(\ki)t - \ki n\rp} + \Ar e^{-i\lp\w(\kr)t - \kr n \rp}
    \\ \label{eq:a_n_plus}
    a_{n_{+}} = \At e^{-i\lp\w(\kt)t - \kt n \rp}
\end{gather}

We assume a continuity-like boundary condition at the defect site:
\begin{gather}
    a_{n_{-} = 0} = a_{n_{+} = 0} \equiv a_{0}
    \\ \label{eq:bc_wave_explicit}
    e^{-i\w(\ki)t} + \Ar e^{-i\w(\kr)t} = \At e^{-i\w(\kt)t} 
\end{gather}
For eqn. \ref{eq:bc_wave_explicit} to hold at all times, we set $\w(\ki) = \w(\kr) = \w(\kt) \equiv \w$ so that $1 + \Ar = \At$. The corresponding relation between $k$'s depends on the dispersion; we consider uniform nearest-neighbor coupling with $\w(k) = 2|J| \cos(k + \thJ)$ as it is most relevant to this work. We then have $\ki + \thJ = \pm (\kr + \thJ) = \pm (\kt + \thJ)$, where any permutation of the signs is valid. The assumption of reflected and transmitted waves implies group velocities that are respectively antiparallel and parallel to the incident wave, so we take
\begin{equation}
    \ki + \thJ = \kt + \thJ = - (\kr + \thJ)
\end{equation}

Considering the equation of motion for $a_{0}$ at the frequency $\w = \w(\ki)$ and substituting (\ref{eq:a_n_minus}, \ref{eq:a_n_plus}):
\begin{gather}
    \dot{a}_{0} = -i \lp \Delta a_{0} + J a_{-1} + J^* a_{1} \rp
    \\
    -i\w a_{0} \ofw = -i \lp \Delta a_{0} \ofw + J a_{-1}\ofw + J^* a_{1} \ofw \rp
\end{gather}
\begin{multline} \label{eq:eom_ao}
    -iJe^{j\ki} - \Ar(iJe^{-i\kr}) 
    \\
    + \At(-i(\Delta - \w) + J^* e^{i\ki}) = 0
\end{multline}
Combining eqns. \ref{eq:eom_ao} and \ref{eq:bc_wave_explicit} and solving, noting that the resulting determinant is typically nonzero:
\begin{equation}
    \begin{pmatrix}
 -iJe^{j\kr} &  -i(\Delta - \w) - iJ^*e^{i\ki} \\ 
 1 & -1 
\end{pmatrix}
\begin{pmatrix}
 \Ar  \\ 
 \At 
\end{pmatrix}
=
\begin{pmatrix}
 iJe^{-i\ki}  \\ 
 -1 
\end{pmatrix}
\end{equation}

\begin{widetext}
\begin{equation} \label{eq:elastic_soln_1}
    \begin{pmatrix}
 \Ar  \\ 
 \At 
\end{pmatrix}
= \frac{1}{\Delta - \w + Je^{-i\kr} + J^{*} e^{i\ki}}
\begin{pmatrix}
 -(\Delta-\w) - Je^{-i\ki} - J^{*}e^{i\ki}  \\ 
 - Je^{-i\ki} + Je^{-i\kr}
\end{pmatrix}
\end{equation}
\end{widetext}

To check the consistency of the postulated solution, consider the equation of motion for a site outside the defect, e.g. $a_{1}$, and substitute the assumed forms:
\begin{multline}
    \dot{a}_{1} = \partial_{t} \lp \At e^{-i\lp\w t - \ki \rp} \rp = -i\w\At e^{-i\lp\w t - \ki \rp}
    \\
     = -i(J a_{0} + J^{*}a_{2}) = -i\At \lp J e^{-i\w t} + J^{*}e^{-i( \w t -2\ki)} \rp
\end{multline}

Here eqn. \ref{eq:a_n_minus} has been used to write $\dot{a}_{1}$ in terms of the transmitted amplitude. Rearranging and using $J = |J|e^{-i\thJ}$ obtains:
\begin{equation} \label{eq:a1_condition}
   -i \At e^{i \ki} \lp \w - 2|J| \cos(\ki + \thJ) \rp = 0
\end{equation}

which indicates the postulated solution satisfies the correct evolution for $a_{1}$. Noting that $\w(\ki) = Je^{-i\ki} + J^{*}e^{i\ki}$ we can simplify eqn. \ref{eq:elastic_soln_1}:
\begin{equation} \label{eq:elastic_soln_2}
\begin{pmatrix}
 \Ar  \\ 
 \At 
\end{pmatrix}
= \frac{1}{\Delta - i v_{g}(\ki)}
\begin{pmatrix}
  - \Delta  \\ 
  -i v_{g}(\ki)
\end{pmatrix}
\end{equation}

where $v_{g}(\ki) = \partial_{\ki}\w = -2|J|\sin(\ki + \thJ)$. This equation provides a clear picture of the scattering: the reflected amplitude arises from the detuning, and the transmitted amplitude is determined by the ratio of group velocity to detuning. Eqn. \ref{eq:elastic_soln_2} is invalid only for the degenerate case with zero detuning and group velocity, where no spatial defect or propagation occurs. The magnitude-squared of both amplitudes are plotted in Fig. \ref{fig:freq_defect}b for representative values of $k_\text{eff} \equiv \ki + \thJ$; the transmitted amplitude approaches 1 symmetrically as $|\Delta/J| \rightarrow 0$, with FWHM increasing from $0$ at $k_\text{eff} = 0$ to $4$ (zero group velocity) at $|k_\text{eff}| = 0.5\pi$ (maximum group velocity).

\end{document}